\documentclass[%
 reprint,
superscriptaddress,
 amsmath,amssymb,
 aps,
]{revtex4-1}

\usepackage{graphicx,subfigure}
\usepackage{dcolumn}
\usepackage{bm}
\graphicspath{ {images/} }
\usepackage{amsmath}
\usepackage{bbm}

\newcommand{\bP}[1]{{\mathbb{P}}\left[{#1}\right]}
\newcommand{\bE}[1]{{\mathbb{E}}\left[{#1}\right]}

\newcommand{\1}[1]{{\bf 1}\left[#1\right]}
\usepackage{mathtools}
\usepackage{amssymb}
\usepackage{float}
\usepackage{enumitem}
\usepackage{mwe}
\usepackage[usenames, dvipsnames]{color}
\usepackage[toc,page]{appendix}
\usepackage{natbib}
\begin{document}

\preprint{APS/123-QED}

\title{Cascading failures in interdependent systems under a flow redistribution model}

\author{Yingrui Zhuang}
\affiliation{
 Department of ECE, Carnegie Mellon University, Pittsburgh, PA 15213 USA
}
\author{Alex Arenas}
\affiliation{
Departament d'Enginyeria Inform\'atica i Matem\'atiques, Universitat Rovira i Virgili, 43007 Tarragona, Spain
}
\author{Osman Ya\u{g}an}
\email{oyagan@ece.cmu.edu}
\affiliation{
 Department of ECE, Carnegie Mellon University, Pittsburgh, PA 15213 USA
}






\begin{abstract}
Robustness and cascading failures in {\em interdependent} systems has been an active research field in the past decade. However, most existing works use percolation-based models where only the largest component of each  network remains functional throughout the cascade. Although suitable for communication networks, this assumption fails to capture the dependencies in systems carrying a flow (e.g., power systems, road transportation networks), where cascading failures are often triggered by {\em redistribution} of flows leading to {\em overloading} of lines.
Here, we consider a model consisting of systems $A$ and $B$ with initial line loads and {\em capacities} given by $\{L_{A,i},C_{A,i}\}_{i=1}^{n}$ and $\{L_{B,i},C_{B,i}\}_{i=1}^{n}$, respectively. When a line fails in system $A$,
 $a$-fraction of its load is redistributed to alive lines in $B$, while remaining $(1-a)$-fraction is redistributed
  equally among all functional lines in $A$; a line failure in $B$ is treated similarly with $b$ giving the fraction to be redistributed to $A$.
We give a thorough analysis of cascading failures of this model initiated by a random attack targeting $p_1$-fraction of lines in $A$ and $p_2$-fraction in $B$. We show that 
(i) the model captures the real-world phenomenon of unexpected large scale cascades and exhibits interesting transition behavior: the final collapse is always first-order, but it can be preceded by a sequence of first and second-order transitions;
(ii) network robustness tightly depends on the coupling coefficients $a$ and $b$, and robustness is maximized  at  non-trivial $a,b$ values in general; (iii) unlike existing models,
interdependence has a multi-faceted impact on system robustness in that interdependency can lead to an {improved} robustness for each individual network. 
\end{abstract}

\pacs{Valid PACS appear here}
\maketitle

\section{Introduction}

With the development of modern technology, networks emerge as the new form of how things work in every aspect of our life, from online social media to cyber-physical systems, from intelligent highways to aerospace systems. Soon we will expect computing and communication capabilities to be embedded in all physical objects and structures and more complex networks to appear \cite{rajkumar2010cyber}. Recently, researchers have become increasingly aware of the fact that most systems do not live in isolation, and that  they exhibit significant {\em inter-dependencies} with each other. In particular, it has been shown that interdependence and coupling among networks lead to dramatic changes in network dynamics, with studies focusing on cascading failure and robustness \cite{li2012cascading,buldyrev2010catastrophic,gao2011robustness,yagan2012optimal,brummitt2012suppressing,rinaldi2004modeling,multiple_percolation}, information and influence propagation \cite{Yagan_realtime,zhuang2016information,YaganGligor,YaganJSAC,YongTNSE}, percolation \cite{parshani2010interdependent,son2012percolation,min2014network,lee2014threshold,wu2014multiple}, etc.

One of the most widely studied network dynamics is the cascade (or, spread) of failures. Due to the coupling between diverse infrastructures such as water supply, transportation, fuel and power stations,  interdependent networks are tend to be extremely  vulnerable \cite{vespignani2010complex}, because the failure of a small fraction of nodes from one network can produce an iterative cascade of failures in several interdependent networks. Blackouts are typical examples of cascading failures catalyzed by the dependencies between networks: the September 28, 2003 blackout in Italy resulted in a widespread failure of the railway network, health care systems, and financial services and, in addition, severely influenced communication networks. As a result, the partial failure of the communication system in turn further impaired the power grid management system. 

Robustness of interdependent networks has  been an active research field after the seminal paper of Buldyrev et al. \cite{buldyrev2010catastrophic}, with the key result being interdependent networks are more vulnerable than their isolated counterparts. However, existing works on cascading failures in interdependent networks focus extensively on percolation-based models  \cite{buldyrev2010catastrophic,parshani2010interdependent,buldyrev2011interdependent,gao2012networks,radicchi2015percolation,di2016cascading}, where a node can function only if it belongs to the largest connected (i.e., giant) component of its own network
; nodes that lose their connection to this giant core are deemed {\em non-functional}. While such models are suitable for communication networks, they fail to accurately capture the dynamics of cascading failures in many real-world systems that are tasked with transporting physical commodities; e.g.,
power networks, traffic networks, etc. In such flow networks, failure of nodes (or, lines) lead to {\em redistribution} of their load to functional nodes, potentially {\em overloading} and failing them. As a result, the dynamics of failures is governed primarily by load redistribution rather than the structural changes in the network. A real-world example to this phenomenon took place on July 21, 2012, when a heavy rain shut down a metro line and caused 100 bus routes to detour, dump stop, or stop operation completely in Beijing \cite{huang2015cascading}. 

\begin{figure}[!t]
\centering
\hspace{-5mm}\includegraphics[width=0.52\textwidth]{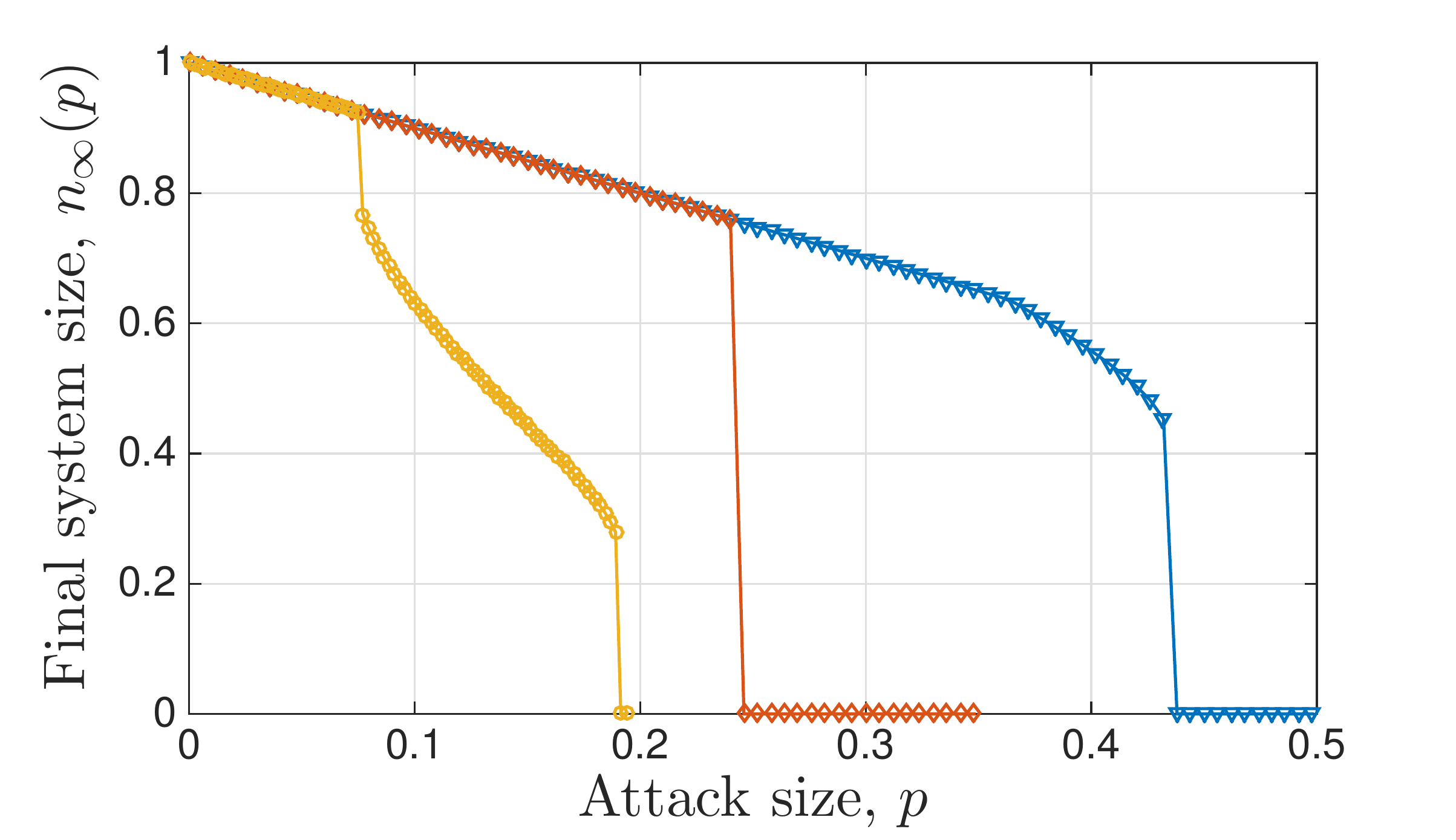}
\caption{\sl Possible transition behaviors under the load redistribution based cascade model. We see that final system collapse is always first order, which may be preceded with one or more  first- or, second-order  transitions. }
\label{fig:single_trans}
\end{figure}

In this paper, we initiate a study on robustness of interdependent networks under a load redistribution based cascading failure model. Our approach is inspired by the fiber-bundle model that has been extensively used to investigate the fracture and breakdown of a broad class of disordered systems; e.g., magnets driven by an applied field \cite{da1999introduction},  earthquakes \cite{moreno2001model,turcotte2004damage}, power system failure \cite{yingrui2016cascadingfailure}, social phenomena \cite{pradhan2010failure}. This model has already been demonstrated to exhibit {\em rich} transition behavior in a single network setting under random attacks of varying size, while being able to capture some key characteristics of real-world cascades \cite{yaugan2015robustness,yingrui2016cascadingfailure}; e.g., see Figure \ref{fig:single_trans}. In particular,
it was shown that the transition point where the system has a total breakdown is always discontinuous, reminiscent of the real-world phenomena of unexpected large-scale system collapses; i.e., cases where seemingly identical attacks leading to entirely different consequences. While this breakdown can take place abruptly without any indicators at smaller attack sizes (as in the middle curve in Figure \ref{fig:single_trans}), it may also be preceded with one or more first-order or second-order transitions (as seen in the other two curves of Figure \ref{fig:single_trans}) that can be taken as early warning signs of a catastrophic cascade. 

We extend the fiber-bundle-like cascading failure model to interdependent networks as follows. Assume that the system consists of $n$ {\em coupled} networks each with a given number of transmission {\em lines}. Every line is given an initial load $L$ and a capacity $C$ defined as the maximum load it can tolerate; if the load on the line exceeds its capacity (for any reason) the line is assumed to fail. The main ingredient of the model is the load redistribution rule: upon failure of a line in any network, the load it was carrying before the failure will be redistributed among all networks in the system, with the proportion received by each network being determined by the {\em coupling coefficients} across networks; see Section \ref{sec:model} for precise details. 
Within each network, we adopt the fiber-bundle-like model \cite{yaugan2015robustness,yingrui2016cascadingfailure} and distribute this received load equally among all {\em functional} lines.

We give a thorough analysis of cascading failures (based on the model described above) in a system of two interdependent networks initiated by a random attack. 
We show that in addition to providing a more realistic model of cascading failures for interdependent systems (as compared to percolation-based models), the model described above gives rise to interesting and novel transition behavior, and challenges the widely accepted notion that interdependence (or, coupling, or, inter-connectivity) is always detrimental for system robustness. 
In particular, we show that 
(i) the model captures the real-world phenomenon of unexpected large scale cascades: final collapse is always first-order, but it can be preceded by a sequence of first and second-order transitions; to the best of our knowledge such behavior has not been observed before in any model.
(ii) network robustness tightly depends on the coupling coefficients and robustness is maximized  at  non-trivial coupling levels in general; (iii) unlike existing models,
interdependence has a multi-faceted impact on system robustness in that interdependency can lead to an improved robustness for each individual network.

We reiterate that although extensive, the literature on cascading failures in interdependent networks is limited to percolation-based models that fail to capture many real-world settings. Load redistribution models on the other hand have mostly been constrained to single-network settings; e.g., \cite{MotterLai,wang2008universal,mirzasoleiman2011cascaded}.
 
The closest work to our paper is by Brummitt et al. \cite{brummitt2012suppressing} where a {\em sandpile} model was studied for two inter-connected networks (each being a random regular graph). Although a similar observation regarding the impact of inter-connectivity was made (that it can sometimes help improve robustness), their work is limited to cascades triggered by increased initial load  on the system (imitating the sand dropping process) instead of random failures or attacks considered here; as such, \cite{brummitt2012suppressing} does not provide any insight regarding the transition behavior of the system against attacks and how that behavior is affected by the level of inter-connectivity \footnote{In addition,  \cite{brummitt2012suppressing} considers a specific load-capacity relation, while our work covers more general settings.}. To the best of our knowledge, the only other relevant work is by Scala et. al.\cite{scala2016cascades} who studied cascades in coupled distribution grids, but again under a load growth model instead of external attacks. 


%
%
%
%
%
%
  

The rest of the paper is organized as follows: we formally define the load redistribution model and analysis tools used in Section \ref{sec:model}. Our analytic results are presented in Section \ref{sec:analytical}, including the solutions for steady-state system sizes. Numerical results are given in section \ref{sec:result}, and we conclude our work in section \ref{sec:conclusion}.

\section{Model Definitions}
\label{sec:model}
We consider a system composed of $n$ networks that interact with each other. Let $\mathcal{N}=\{1,\ldots,n\}$ denote the set of all networks in the system. For each $i \in \mathcal{N}$, we assume that network $i$ has $N_i$ lines $\mathcal{L}_{1,i}, \ldots, \mathcal{L}_{N_i,i}$ 
with initial loads $L_{1,i}, \ldots, L_{N_i,i}$. Each of these lines is associated with a capacity $C_{1,i}, \ldots, C_{N_i,i}$ above which the line will be tripped. In other words, $C_{k,i}$ defines the maximum flow that line $k$ in network $i$ can sustain and is given by
\begin{equation}
C_{k,i} = L_{k,i} + S_{k,i}, \qquad i \in \mathcal{N},\quad k = 1, \ldots, N_i
\nonumber
\end{equation}
where $S_{k,i}$ denotes the free space on line $k$ in network $i$, i.e., the maximum amount of extra load it can take. The load-free space pairs $\{L_{k,i},S_{k,i}\}_{k=1}^{N_i}$ are independently and identically distributed with 
\[
P_{L_iS_i}(x,y):=\bP{L_{k,i} \leq x,~ S_{k,i} \leq y}, \quad k=1, \ldots, N_i
\]
for each  $i \in \mathcal{N}$. The corresponding joint probability density function is given by $p_{L_iS_i}(x,y)=\frac{\partial^2}{\partial x \partial y} P_{L_iS_i}(x,y)$. In order to avoid trivial cases, we assume that $S_{k,i}>0$ and $L_{k,i}>0$ with probability one for each $i \in \mathcal{N}$ and each $k=1, \ldots, N_i$. 
Finally, we assume that the marginal densities $p_{L_i}(x)$ and $p_{S_i}(y)$ are continuous on their support. 

Initially, $p_i$-fraction of lines are attacked (or failed) randomly in network $i$, where $p_i \in [0,1]$. The load on failed lines will be redistributed within the original network and/or shed to other coupled networks depending on the underlying redistribution rules governing the system. Further failures may then take place within the initially attacked network or in the coupled ones due to lines undertaking extra load exceeding their capacity; this in turn leads to further redistribution in all constituent network, potentially leading to a {\em cascade} of failures. The cascade of failures taking place simultaneously within and across networks leads to an interesting dynamical behavior and an intricate relationship between the level of coupling and the system's overall robustness.

The cascade process is monotone (once failed, a line remains so forever), and thus it will eventually {\em stop}, potentially when all lines in all networks have failed. Otherwise a positive fraction of lines may survive the cascade in one or more of the constituent networks. One of our main goals in this paper is to characterize the fraction of alive lines in each network at that \lq steady state'; i.e., at the point where cascades stop. 
To that end, we provide a {\em mean-field} analysis of dynamical process of cascading failures. Under this approach, it is assumed that when a line fails, its flow will be redistributed to its own network as well as to other networks with the proportion redistributed to each network determined by {\em coupling coefficients} among the networks; more on this later. Each network will then distribute its own share of the failed load {\em equally and globally} among all of its remaining lines. 

Although simple, the equal load redistribution model is able to capture the long-range nature of failure propagation in physical systems (e.g., Kirchhoff's law for power networks), at least in the mean-field sense, as opposed to the topological models \cite{crucitti2004model,wang2008universal} where failed load is redistributed only locally among neighboring lines. In our case, it also enables focusing on how coupling and interdependence of two arbitrary networks affect their overall robustness, even if individual network topologies might be unknown.


As mentioned before, the flow of a failed line in a network will not only be redistributed internally, but will also be shed to other coupled networks. The proportion of load to be shed from a failed line in network $i$ to network $j$ is determined by the coupling coefficient $a_{ij}$, where we have $\sum_{j \in \mathcal{N}}a_{ij}=1$ for all $i$ in $\mathcal{N}$; thus, $1-\sum_{j \in \mathcal{N}-\{i\}}a_{ij}$ gives the fraction of the load that will be redistributed internally in network $i$. The load received in each network is then shared equally among all of its functional lines. Upon redistribution of flows, the load on each alive line will be updated potentially leading to some lines having more load than their capacity, and thus failing. Subsequently, the load of those additionally failed lines will be redistributed in the same manner, which in turn may cause further failures, possibly leading to a cascade of failures in both the initiating networks and their coupled networks. This phenomenon imitates the interdependent systems in real world where the failure in one network, such as power network, can affect the behavior of another network, such as water system and financial systems.

\begin{figure}[t]
\centering
\includegraphics[width=0.38\textwidth]{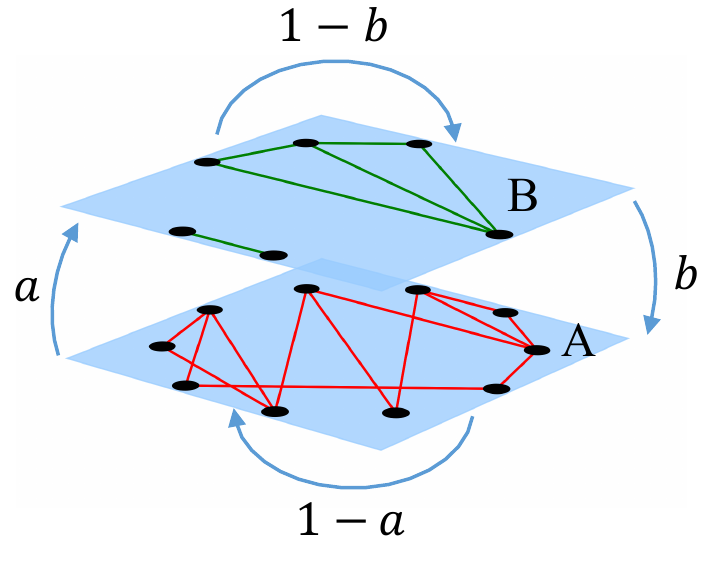}
\vspace{-2mm}
\caption{\label{fig:sys} \sl Illustration of a two-network system. When failures happen in network $B$, $b$-portion of the failed loads goes to network $A$ and $(1-b)$-portion stays in $B$. Similarly in network $A$, $(1-a)$-portion stays and $a$-portion goes to $B$. Failed loads will be redistributed equally and globally among the remaining lines in each network.
}
\end{figure}

For the ease of exposition, we consider a two-network system in the rest of the paper, although our results can be extended trivially to arbitrary number of networks. Consider a system composed of networks $A$ and $B$ that are {\em interdependent} in the following manner \footnote{Of course, there are other ways for two networks to be  {\lq\lq interdependent"} with each other. Here, we use this term with its general meaning, i.e., that failures in one network may lead to failures in the other and vice versa, potentially leading to a cascade of failures. Our model constitutes a special case where interdependence emerges from the {\em inter-connectivity} between the two networks}: when a failure happens in network $A$, $a$ fraction of the failed load is transferred to network $B$, while the remaining $1-a$ fraction being redistributed internally in $A$. Similarly upon failures in network $B$, $b$ fraction of the failed load will be shed to network $A$; here $a,b \in [0,1]$ are system defined constants. An illustration of the system can be found in Figure \ref{fig:sys}. We assume that initially $p_1$-fraction of lines in network $A$ and $p_2$-fraction of lines in network $B$ fail randomly. The initial attacks may cause cascading failures, and if one of the network collapses (i.e., if all of its lines fail)  during this process, the other network will take over the rest of the load in it and function as a single network from that point on. 


With appropriate meanings of load and capacity, this type of load oriented models can capture the dependencies in a wide range of physical systems; e.g., two smart-grid operators coupled to provide better service \cite{nguyen2013detecting}, two banks highly correlated for collective risk shifting \cite{elsinger2006risk}, or two interacting transportation networks \cite{gao2012networks}. In what follows, we provide an analytic solution for the dynamics of cascading failures 
in the model described above.


\section{Analytic Results}
\label{sec:analytical}
We now provide the mean-field analysis of cascading failures in the two-network interdependent system. Without loss of generality, we assume that both networks have the same number of lines, i.e., $N_A = N_B = N$. We assume that time is divided into discrete steps, $t=1, 2, \ldots$. For each time stage $t$, and with $X \in \{A,B\}$, we use the following notation:

$f_{t,X}$: fraction of failed lines until $t$;

$F_{t,X}$: total load from lines that fail exactly at time $t$  within network $X$;

$Q_{t,X}$: extra load to be redistribution at $t$ per alive line in $X$;  

$N_{t,X}$: number of alive lines at $t$ in $X$ before redistribution.

In what follows, we occasionally provide expressions only for the quantities regarding network $A$, while the corresponding expressions for network $B$ (that are omitted in the text for brevity) can be obtained similarly.

Initially, $p_1$-fraction of lines in network $A$ and $p_2$-fraction of lines in network $B$ are attacked (or failed) randomly. Thus, the fraction of failed lines within each network at $t=0$ is given by
$$f_{0,A} = p_1, \quad f_{0,B} = p_2$$,
while the number of alive lines satisfy
\begin{align}\nonumber
	N_{0,A} &= (1 - f_{0,A})N = (1 - p_1)N
	\\ \nonumber
	N_{0,B} &= (1 - f_{0,B})N = (1 - p_2)N
	\end{align}
Because the initially attacked lines are selected uniformly at random, the total load from failed lines (in the mean-field sense) satisfy
	\begin{align}\nonumber
	F_{0,A} &= \bE{L_A} \cdot f_{0,A} \cdot N = \bE{L_A} \cdot p_1 \cdot N
	\\ \nonumber
	F_{0,B} &= \bE{L_B} \cdot f_{0,B} \cdot N = \bE{L_B} \cdot p_2 \cdot N
	\end{align}
    
Based on the equal redistribution rule and the load shedding rule between the two interdependent networks, the extra load per alive line in network $A$ at $t=0$ is:

\begin{align*}
Q_{0,A} &=  \frac{(1-a) \cdot F_{0,A}   +  b \cdot F_{0,B} }{(1 - f_{0,A})N} \\
&= \frac{(1-a) \cdot \bE{L_A} \cdot p_1  + b \cdot \bE{L_B} \cdot p_2 }{1 - p_1}
\end{align*}

and similarly for network $B$:
$$Q_{0,B} = \frac{a  \cdot \bE{L_A} \cdot p_1  + (1-b) \cdot \bE{L_B} \cdot p_2 }{1 - p_2}$$

At stage $t=1$, line $k$ in network $A$ that survives the initial attack will fail if and only if the updated loads exceed its capacity, i.e., if $L_{k,A} + Q_{0,A} \geq L_{k,A} + S_{k,A}$, or equivalently, if $S_{k,A} \leq Q_{0,A}$. Based on this condition, the fraction of failed lines at $t=1$ is given by
\begin{align*}
f_{1,A} &= f_{0,A} + (1-f_{0,A}) \cdot \bP{S_{A} \leq Q_{0,A}} \\
&= 1 - (1 - f_{0,A})\bP{S_{A} > Q_{0,A}}
\end{align*}

To compute the extra load per alive line in each network at $t=1$, we need to know the lines that fail exactly at this stage in each network (so that their load can be appropriately redistributed to both networks according to the coupling coefficients). Namely, we need to find the lines that survive the initial attack, but have smaller free space than the redistributed load $Q_{0,A}$ or $Q_{0,B}$ from the previous stage. Let $\mathcal{A}$ and $\mathcal{B}$ be the initial set of lines that are attacked or failed initially in network $A$ and $B$, respectively. Then, the total load on these failed lines in network $A$ at $t=1$ can be derived as



\begin{align}
F_{1,A}&= \bE{\sum_{i \notin \mathcal{A},S_{i,A} \leq Q_{0,A}}(L_{i,A}+ Q_{0,A})} \nonumber \\
&= \bE{\sum_{i \notin \mathcal{A}}(L_{i,A}+ Q_{0,A}) \cdot \1{S_{i,A} \leq Q_{0,A}}} \nonumber \\
&= (1-p_1)N\bE{(L_{A}+ Q_{0,A}) \cdot \1{S_A \leq Q_{0,A}}}
\nonumber
\end{align}
where $\1{\cdot}$ is the indicator function \footnote{Let $E$ be an event. Then, $\1{E}$ is a Binomial random variable that takes the value of $1$ if $E$ takes place, and $0$ otherwise}; here we used the fact that for each line $i$ in $A$, $L_i,S_i$ follow the same distribution $p_{L_A,S_A}$. Similarly for network $B$, we have
\begin{align}
F_{1,B} &= \bE{\sum_{i \notin \mathcal{B},S_{i,B} \leq Q_{0,B}}(L_{i,B}+ Q_{0,B})} \nonumber \\
&= (1-p_2)N\bE{(L_{B}+ Q_{0,B}) \cdot \1{S_B \leq Q_{0,B}}}
\nonumber
\end{align}

The load of these lines failed at stage 1 will then be redistributed internally and across network, based on the aforementioned coupling coefficients. This leads to the extra load per alive line in network $A$ at $t=1$ being given by
\begin{align}
&Q_{1,A} 
\nonumber \\
&= Q_{0,A} + \frac{(1-a) \cdot F_{1,A} + b \cdot F_{1,B}}{N(1-f_{1,A})} \nonumber \\
&= Q_{0,A} + \nonumber\\ 
& \quad  \dfrac{ \splitdfrac{(1-a)(1-p_1)  \bE{(L_{A}+Q_{0,A}) \cdot \1{S_A \leq Q_{0,A}} }  }{ + b(1-p_2)  \bE{(L_{B}+Q_{0,B}) \cdot \1{S_B \leq Q_{0,B}}} }}{1-f_{1,A}} 
\nonumber
\end{align}
$Q_{1,B}$ can be written in a similar manner.

At $t=2$, more lines will fail because of the redistribution in the previous stage. The condition for a line to fail exactly at $t=2$ is: (i) it doesn't belong to the initial attack set $\{\mathcal{A}$, $\mathcal{B}\}$; (ii) it survived the redistribution in the previous stage $t=1$; and (iii) its capacity is less than the updated total load after redistribution at $t=2$. From this we can derive the fraction of failed lines till $t=2$ as
\begin{align}
&f_{2,A} = 1-(1-f_{1,A})\bP{S_A > Q_{1,A} ~|~ S_A > Q_{0,A}} \nonumber \\
&f_{2,B} = 1-(1-f_{1,B})\bP{S_B > Q_{1,B} ~|~ S_B > Q_{0,B}}\nonumber
\end{align}

Then, the total load from lines that fail exactly at $t=2$ in network $A$ is given by
\begin{align}
&F_{2,A} 
\nonumber \\
&= \bE{\sum_{i \notin \mathcal{A},Q_{0,A}<S_{i,A} \leq Q_{1,A}}(L_{i,A}+ Q_{1,A})} \nonumber \\
&= (1-p_1)N \bE{(L_{A}+ Q_{1,A})  \1{Q_{0,A}<S_A \leq Q_{1,A}}} \nonumber
\end{align}
Similarly in network $B$, we have
\begin{align}
& F_{2,B} 
\nonumber \\
&= \bE{\sum_{i \notin \mathcal{B},Q_{0,B}<S_{i,B} \leq Q_{1,B}}(L_{i,B}+ Q_{1,B})} \nonumber \\
&= (1-p_2)N \bE{(L_{B}+ Q_{1,B})  \1{Q_{0,B}<S_B \leq Q_{1,B}}}
\nonumber
\end{align}

With the total loads on failed lines $F_{2,A}$, $F_{2,B}$ and the fraction of failed lines $f_{2,A}$, $f_{2,B}$ in each network, the extra load per alive line in network $A$ at stage $t=2$ can be calculated as 
\begin{align}
&Q_{2,A} 
\nonumber \\
& = Q_{1,A} + \frac{(1-a)F_{2,A} + bF_{2,B}}{N(1-f_{2,A})} \nonumber \\
&= Q_{1,A} +\nonumber \\
&\dfrac{ \splitdfrac{(1-a)(1-p_1)  \bE{(L_{A}+Q_{1,A}) \cdot \1{Q_{0,A}<S_A \leq Q_{1,A}} } }{ + b(1-p_2) \bE{(L_{B}+Q_{1,B}) \cdot \1{Q_{0,B}<S_B \leq Q_{1,B}}} }}{1-f_{2,A}}
\nonumber
\end{align}
A similar expression gives $Q_{2,B}$. 

In light of the above derivation, the form of the recursive equations is now clear: for each time stage $t=0,1,\ldots,$ we have
\begin{widetext}
\begin{align}
&f_{t+1,A} = 1-(1-f_{t,A})\bP{S_A > Q_{t,A} ~|~ S_A > Q_{t-1,A}} \nonumber \\
\nonumber \\
&N_{t+1,A} = (1-f_{t+1,A})N
\nonumber \\
\label{eq:recursive} \\
&Q_{t+1,A} = Q_{t,A} + \dfrac{ \splitdfrac{(1-a)(1-p_1) \bE{(L_{A}+Q_{t,A})\cdot \1{Q_{t-1,A}<S_A \leq Q_{t,A}} }}{+ b(1-p_2)  \bE{(L_{B}+Q_{t,B}) \cdot \1{Q_{t-1,B}<S_B \leq Q_{t,B}}}  }}{1-f_{t+1,A}}, \nonumber
\end{align}
\end{widetext}
and similarly for network B. 

From (\ref{eq:recursive}) we can see that the cascade of failures will stop and the steady state will be reached only when the number of alive lines doesn't change in both networks, i.e., $N_{t+2,A}=N_{t+1,A}$, $N_{t+2,B}=N_{t+1,B}$. This is equivalent to having
\begin{align}
\label{eq:stop_general}
    &\bP{S_A > Q_{t+1,A} ~|~ S_A > Q_{t,A}} = 1, ~\text{and} \nonumber \\
    &\bP{S_B > Q_{t+1,B} ~|~ S_B > Q_{t,B}} = 1 
\end{align}
In other words, whenever we have finite $Q_{t+1,A}$, $Q_{t,A}$, $Q_{t+1,B}$ and $Q_{t,B}$ values that satisfy (\ref{eq:stop_general}), cascading failures will stop and the system will reach the steady state.

The recursive expressions (\ref{eq:recursive}) can be simplified further in a way that will make computing the final system sizes (i.e., fraction of alive lines at steady-state) much easier.
Firstly, we use the first expression in (\ref{eq:recursive}) repeatedly for each $t=0, 1, \ldots$ to get
\begin{eqnarray}\nonumber
\begin{array}{ll}
1-f_{t+1,A} &= (1-f_{t,A}) \bP{S_A > {Q_{t,A}} ~|~ S_A > {Q_{t-1,A}}} \\
1-f_{t,A} &= (1-f_{t-1,A}) \bP{S_A > {Q_{t-1,A}} ~|~ S_A > {Q_{t-2,A}}} \\
~~~\vdots & \\
1-f_{1,A} &= (1-f_{0,A}) \bP{S_A > Q_{0,A}} 
\end{array}
\end{eqnarray}
Multiplying these equations together, we obtain
\[
1-f_{t+1,A}  = (1-f_{0,A}) \prod_{\ell = 0}^{t} \bP{S_A > {Q_{\ell,A}} ~\big|~ S_A > Q_{\ell-1,A}},
\]
where we set $Q_{-1,A} = 0$ for convenience. Using the fact that $Q_{t,A}$ is non-decreasing in $t$, i.e., $Q_{t+1,A}\geq Q_{t,A}$ for all $t$, we then get
\begin{align}
&1-f_{t+1,A}
\nonumber 
\\
&=  (1-f_{0,A}) 
\nonumber \\ 
&~~~ \cdot \frac{\bP{S_A > Q_{t,A}}}{\bP{S_A > Q_{t-1,A}}} 
\cdots \frac{\bP{S_A > Q_{1,A}}}{\bP{S_A > Q_{0,A}}} \cdot \bP{S_A > Q_{0,A}} \nonumber \\
&= (1-p_1)\bP{S_A > Q_{t,A}} 
  \label{eq:simplified_f_t}
\end{align}
as we recall that $f_{0,A}=p_1$. 

Using the simplified result (\ref{eq:simplified_f_t}) in
(\ref{eq:recursive}), we now get
\begin{widetext}
\begin{align}
&f_{t+1,A} = 1 - (1-p_1)\bP{S_A > Q_{t,A}}  \nonumber \\
\nonumber \\
&N_{t+1,A} = (1-p_1)\bP{S_A > Q_{t,A}} N
\nonumber  \\
\label{eq:recursive_2}
 \\
&Q_{t+1,A} = Q_{t,A} +  \dfrac{ \splitdfrac{(1-a)(1-p_1) \bE{(L_{A}+Q_{t,A})\cdot \1{Q_{t-1,A}<S_A \leq Q_{t,A}} } }{ + b(1-p_2) \bE{(L_{B}+Q_{t,B}) \cdot \1{Q_{t-1,B}<S_B \leq Q_{t,B}}} }}{(1-p_1)\bP{S_A > Q_{t,A}} }
\nonumber
\end{align}
\end{widetext}
leading to a much more intuitive expression than before. To see why (\ref{eq:recursive_2}) makes sense realize that for a line to survive stage $t+1$ without failing, it is necessary and sufficient that it survives the initial attack (which happens with probability $1-p_1$ for line in network $A$) {\em and} its free-space is greater than the total additional load $Q_{t,A}$ that has been shed on it (which happens with probability $\bP{S_A > Q_{t,A}}$. This explains the first and second expressions in (\ref{eq:recursive_2}).
For the last equation that computes $Q_{t+1,A}$, the extra load per alive line at the end of stage $t+1$ (to be redistributed at stage $t+2$), we  write it as the previous extra load $Q_{t,A}$ plus the extra load from lines that fail {\em precisely} at stage $t+1$. For a line in network $A$, failing precisely at stage $t+1$ implies that the line was not in the initial attack (happens with probability $1-p_1$) {\em and} its free space falls in $(Q_{t-1,A}, Q_{t,A}]$ so that it survived the previous load shedding stage but not the current one. Arguing similarly for lines in network $B$ and recalling the redistribution rule based on coupling coefficients, we can see that the nominator in the second term of $Q_{t+1,A}$ (in  (\ref{eq:recursive_2})) gives the additional new load that will be shed on the alive lines of $A$. The whole expression is now understood upon recalling that $(1-p_1)\bP{S_A > Q_{t,A}}$ gives the fraction of lines from $A$ that survive stage $t+1$ to take this extra load. 


It is now easy to realize that the dynamics of cascading failures is fully governed and understood by the recursions on $Q_{t,A}, Q_{t,B}$ given by 
\begin{widetext}
\begin{align}
\label{eq:recursive_Qa}
&Q_{t+1,A} = Q_{t,A} +  \dfrac{ \splitdfrac{(1-a)(1-p_1) \bE{(L_{A}+Q_{t,A})\cdot \1{Q_{t-1,A}<S_A \leq Q_{t,A}} } }{ + b(1-p_2) \bE{(L_{B}+Q_{t,B}) \cdot \1{Q_{t-1,B}<S_B \leq Q_{t,B}}} }}{(1-p_1)\bP{S_A > Q_{t,A}} }
\\ 
\nonumber \\
&Q_{t+1,B} = Q_{t,B} +  \dfrac{ \splitdfrac{a(1-p_1) \bE{(L_{A}+Q_{t,A})\cdot \1{Q_{t-1,A}<S_A \leq Q_{t,A}} } }{ + (1-b)(1-p_2) \bE{(L_{B}+Q_{t,B}) \cdot \1{Q_{t-1,B}<S_B \leq Q_{t,B}}} }}{(1-p_2)\bP{S_B > Q_{t,B}} }
\label{eq:recursive_Qb}
\end{align}
\end{widetext}
with the conditions for reaching the steady-state still being (\ref{eq:stop_general}). Put differently, in order to find the {\em final} system sizes, we need to iterate (\ref{eq:recursive_Qa})-(\ref{eq:recursive_Qb}) for each $t=0,1, \ldots$ until the stop condition 
(\ref{eq:stop_general}) is satisfied. Let $t^{\star}$ be the stage steady-state is reached and $Q_A^{\star},Q_B^{\star}$ be the corresponding values at that point. 
The final system sizes $n_{\infty,A}$ and $n_{\infty,B}$, defined as the fraction of alive lines in network $A$ and $B$  at the steady state, respectively,  
can then be computed simply from (viz.~(\ref{eq:simplified_f_t}))
\begin{equation} \label{eq:final_size_ab}
\begin{split}
&n_{\infty,A} = 1 - f_{\infty,A} = (1-p_1)\bP{S_A > Q_A^{\star}}  \\
&n_{\infty,B} = 1 - f_{\infty,B} = (1-p_2)\bP{S_B > Q_B^{\star}}. 
\end{split}
\end{equation}

%


The expressions given above for the steady-state of cascading failures in interdependent systems constitute a non-deterministic, nonlinear system of equations, which often do not have to closed-form solution; contrast this with the single network \cite{yingrui2016cascadingfailure} case, where it is possible to provide a closed form solution to the final system size. Therefore, in the interdependent network case, we solve $\{Q_A^{\star}, ~ Q_B^{\star} \}$ by numerically iterating over (\ref{eq:recursive_Qa})-(\ref{eq:recursive_Qb}).
The difficulty of obtaining a closed-form expression for final system sizes arises due to the recursive shedding of load across the two networks. At each stage of the cascade, both networks send a portion of the load from its failed lines to the other network, while receiving a portion of load from the lines failed in the coupled network. Furthermore, the load a line was carrying right before failure depends directly on the extra load per alive line (which decide who fails in the next stage) at the time of its failure. This is why we need to keep track of the set of lines that fail \textit{precisely} at a particular stage to be able to obtain an exact account of these loads \footnote{This is also evident from (\ref{eq:recursive}) where we see that $Q_{t+1}$ depends not only on $Q_t$ but also on $Q_{t-1}$}. As a result, the final system size can only be obtained by running over the iterations and identifying the first stage at which the stop conditions (\ref{eq:stop_general}) are satisfied. 



\bigskip
\section{Numerical Results}
\label{sec:result}

\subsection{Final system size under different load-free space distributions and coupling coefficients}
\label{subsec:general}

To verify our analysis with simulations, we choose different load-free space distributions under various coupling coefficients. Throughout, we consider three commonly used families of distributions: i) Uniform, ii) Pareto, and iii) Weibull. These distributions are chosen here because they cover a wide range of commonly used and representative cases. In particular,  uniform distribution provides an intuitive baseline. Distributions belonging to the Pareto  family are also known as a {\em power-law} distributions and have been observed in many real-world networks including the Internet, the citation network, as well as power systems \cite{pahwa2010topological}. 
Weibull distribution is widely used in engineering problems involving reliability and survival analysis, and contains several classical distributions as special cases; e.g., Exponential, Rayleigh, and Dirac-delta. 
The corresponding probability density functions of these distributions are given below for a generic variable $L$.
\begin{itemize}
\item Uniform Distribution: $L \sim U(L_{\textrm{min}},L_{\textrm{max}})$. 
\[
p_L(x)=\frac{1}{ L_{\textrm{max}}  - L_{\textrm{min}} } \cdot \1{ L_{\textrm{min}} \leq x  \leq L_{\textrm{max}} }
\]
\item Pareto Distribution: $L \sim {Pareto}(L_{\textrm{min}}, \beta)$. With
$L_{\textrm{min}}>0$ and $\beta>0$, the density is given by
\[
p_L(x) =  L_{\textrm{min}}^{\beta} \beta x^{-\beta-1} \1{x \geq L_{\textrm{min}}}.
\]
\item Weibull Distribution: $L \sim Weibull(L_{\textrm{min}},\lambda, k)$. With $\lambda, k, L_{\textrm{min}}>0$, the density is given by
\begin{align}
&p_L(x)
\nonumber \\ \nonumber
&= \frac{k}{\lambda} \left(\frac{x-L_{\textrm{min}}}{\lambda} \right)^{k-1} e^{-\left(\frac{x-L_{\textrm{min}}}{\lambda} \right)^{k}}
 \1{x \geq L_{\textrm{min}}}
 \end{align}
The case $k=1$ corresponds to the exponential distribution, and $k=2$ corresponds to Rayleigh distribution. 
\end{itemize}

\begin{figure}[!t]
        \includegraphics[width=0.5\textwidth]{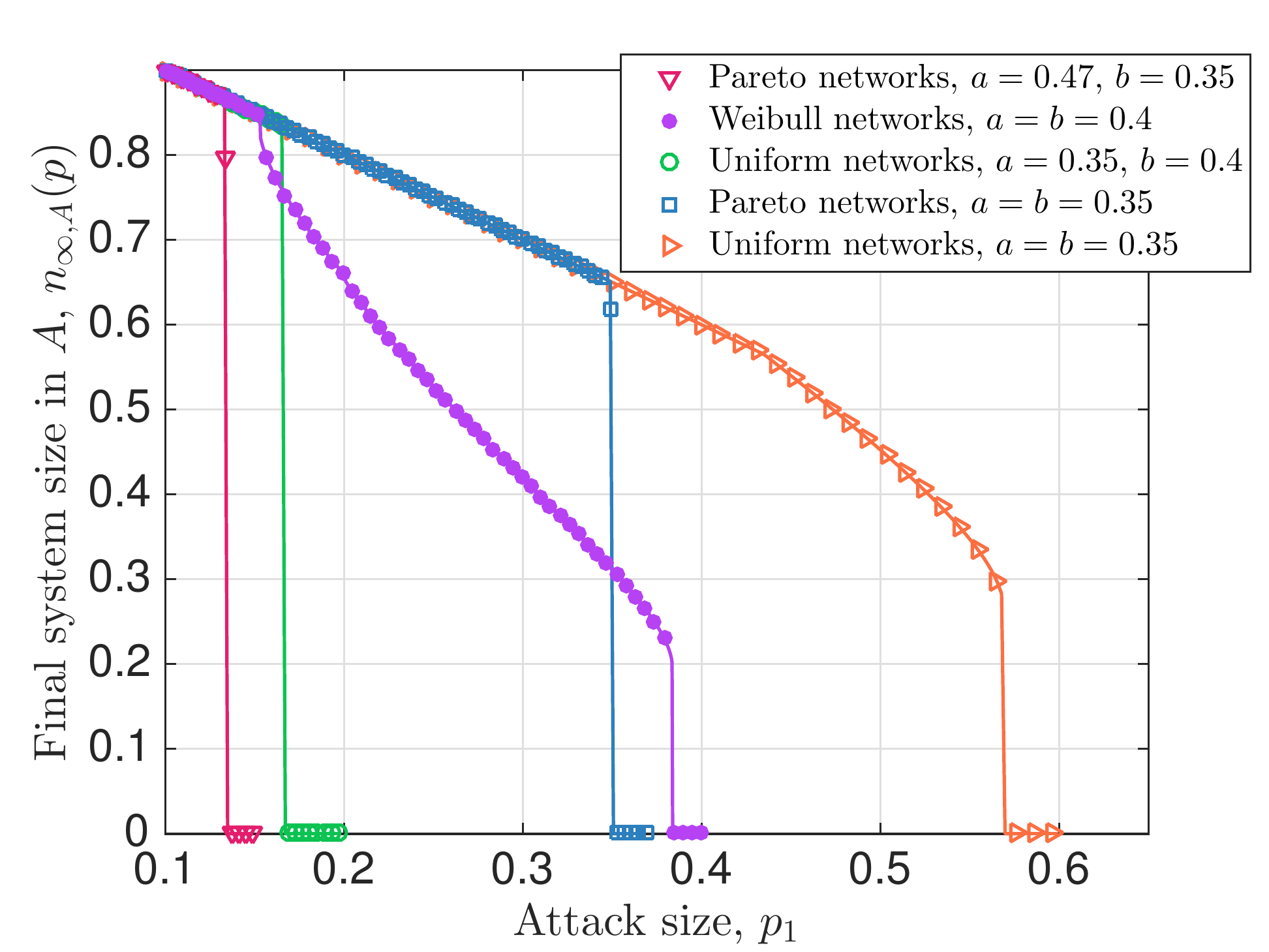}
    \caption{\sl Final system size under different load-free space distributions and coupling coefficients. We observe interesting transition behaviors under different load-free space distributions and coupling level, and the simulation represented in symbol matches with the analytical results represented in lines.
    }
    \label{fig:combine_all}
\end{figure}

In all simulations, we fix the network size at $N=10^7$, and for each set of parameters being considered 
we run 20 independent experiments. The results are shown in Fig.~\ref{fig:combine_all} where symbols represent the empirical value of the final system size $n_{\infty,A}$ of network $A$ (obtained by averaging over 20 independent runs for each data point), and lines represent the analytical results computed from (\ref{eq:final_size_ab}). We see that theoretical results match the simulations very well in all cases. The specific distributions used in Fig. \ref{fig:combine_all} are, from left to right, (i) $L_A \sim Pareto(10,2)$, $S_A=0.7 L_A$, $L_B \sim Pareto(15,1.5)$, $S_B=0.4 L$, and initial attacks are set to $p_2=p_1$; (ii) $L_A, L_B \sim Weibull( 10, 100, k=0.6)$, $S_A=1.74 L_A$,  $S_B=1.5 L_B$, and $p_2=0$; (iii) $L_A \sim U[10,30]$, $S_A \sim U[5,20]$, $L_B \sim U[20,40]$, $S_B \sim U[20,75]$, and $p_1=p_2$; (iv) $L_A,L_B \sim Pareto(10,2)$, $S_A =0.7 L_A$, $S_B= 0.7 L_B$, and $p_1=p_2$; (v) $L_A,L_B \sim U[10,30]$, $S_A,S_B \sim U[10,65]$, and $p_2=0$. 



The plots in Fig. \ref{fig:combine_all} demonstrate the effect of the load-free space distribution as well as coupling level on the robustness of the resulting interdependent system. We see that both the family that the distribution belongs to (e.g., Uniform, Weibull, or Pareto) as well as the specific parameters of the family affect the behavior of $n_{\infty,A}(p)$. For instance, the curves representing the two cases where load and free space in both networks follow a Uniform distribution demonstrate that both abrupt ruptures and ruptures with a preceding divergence are possible in this setting, depending on the parameters. Both cases on Pareto networks give an abrupt breakdown at the final point, and we see that Weibull distribution gives rise to a richer set of possibilities for the transition of final system size $n_{\infty,A}(p)$. Namely, we see that not only we can observe an abrupt rupture, or a rupture with preceding divergence (i.e., a second-order transition followed by a first-order breakdown), it is also possible that $n_{\infty,A}(p)$ goes through a first-order transition (that does not breakdown the system) followed by a second-order transition that is followed by an ultimate first-order breakdown; see the behavior of the purple circled line in Fig. \ref{fig:combine_all}. Thus in the next section, we will use Weibull distribution to explore the interesting transition behaviors observed in interdependent systems composed of two identical networks.

\subsection{Transition behavior for two identical networks}
To explore the effect of coupling and interdependency on the robustness of networks, we couple two (statistically) identical networks. Put differently, we consider networks $A$ and $B$ where the load and capacity of each of their lines are drawn independently from the same distribution. We also assume that they are coupled together in a symmetric way, i.e., that $a=b$. This is a commonly seen case of an interdependent systems where networks of similar characteristics establish a coupling for mutual benefit; e.g., two grid distributors or financial institutions with similar characteristics.  More importantly, this will help us understand the affect of coupling with another identical system on the robustness of a given system; the seminal results of Buldyrev et al. \cite{buldyrev2010catastrophic} suggest that coupling leads to increased vulnerability under percolation based models. 

\begin{figure}[!t]
\hspace{-7mm}
\includegraphics[width=0.5\textwidth]{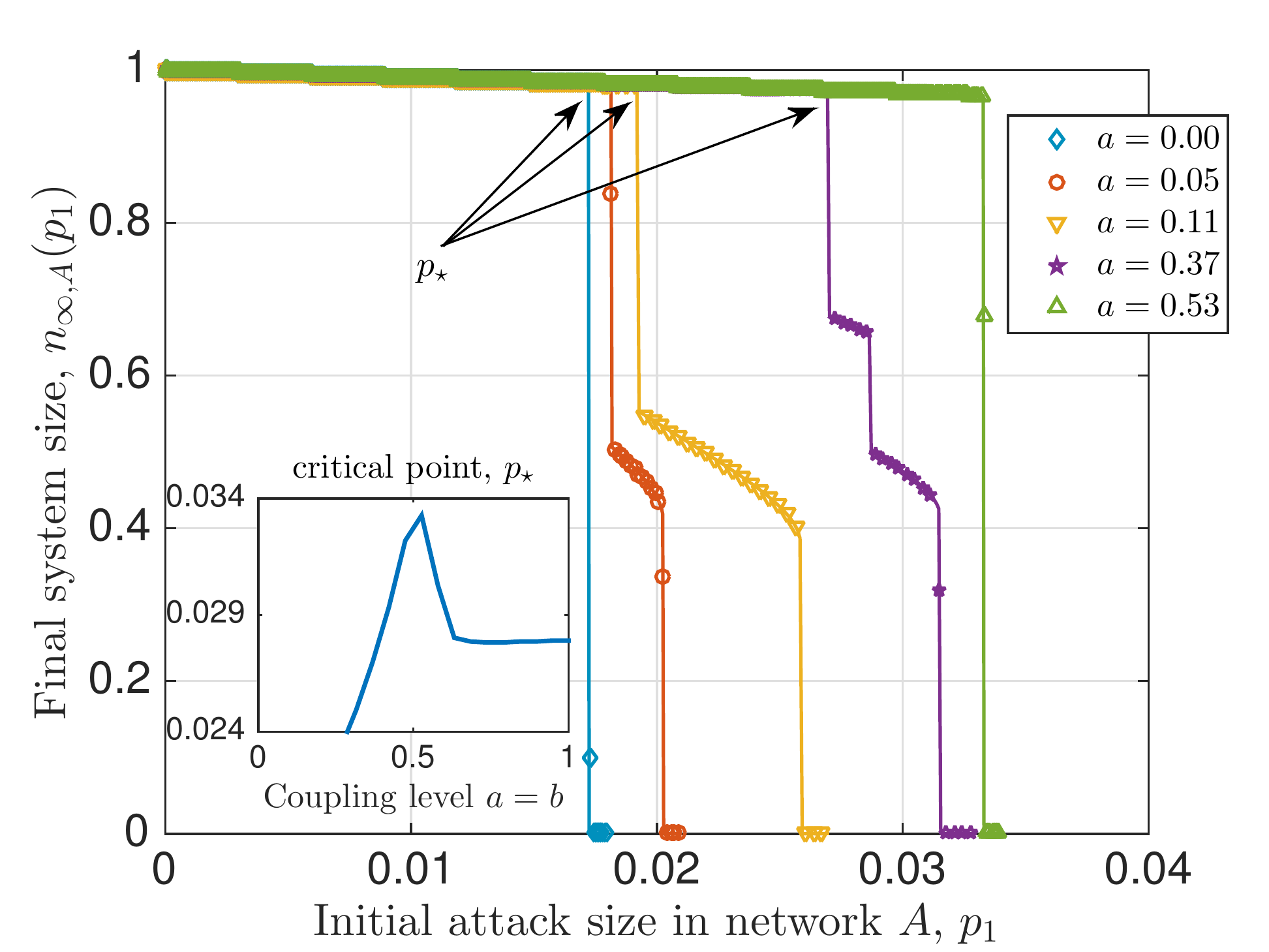}
\caption{\sl Effect of coupling on the robustness of a single system.
We see that contrary to percolation-based models, robustness can indeed be improved by having non-zero coupling between the constituent networks.   
 {\em Inset.} The critical point $p_{\star}$ defined as the smallest $p_1$ at which $n_{\infty,A}(p_1)$ deviates from $1-p_1$.  The optimal (i.e., largest) $p_{\star}$ is attained at a non-trivial coupling level $a = b = \simeq 0.53$.}
\label{fig:2_net_trans}
\end{figure}

With these motivations in mind, we let the initial loads in both networks follow a Weibull distribution, 
with shape parameter $k=0.4$, scale parameter $\lambda = 100$, and minimum initial load $L_{min}=10$. The free space is assigned proportional to the initial load on each line with a tolerance factor $\alpha$, i.e. $S=\alpha L$ where $\alpha=0.6$. The network size is fixed at $N=10^8$. 
We attack $p$-fraction of lines randomly in network $A$, and observe the dynamics of failures driven by the load redistribution across and within the two networks. We then compute the final (i.e., steady-state) size of network $A$ as a function of initial attack sizes $p$  under different values of the coupling coefficient $a$. The results are depicted in Fig. \ref{fig:2_net_trans}, where symbols represent simulation results averaged over 20 independent runs, while lines correspond to our analytical results; in all parameter settings, we observed little to no variance in the final system size across the 20 independent experiments \footnote{We believe this is because the network size $N$ is taken to be very large in the experiments and the random variable $n_{\infty,A}(p_1)$ converges {\em almost surely} to its mean (e.g., by virtue of Strong Law of Large Numbers); though it is beyond the scope of this paper to prove this.}. 

A number of interesting observations can be made from Fig. \ref{fig:2_net_trans}.
First, we see that coupling level can lead to significant changes in the robustness against random attacks. In particular, 
the inset in Fig. \ref{fig:2_net_trans} plots the critical attack size $p_{\star}$ at which the final network size deviates from the $1-p$ line; given attack size $p$, the final system size can be at most $1-p$, which happens when the initial attack does {\em not} lead to any further failures. The network can be deemed to be more robust when $p_{\star}$ is larger. {\em An interesting observation is that unlike the traditional percolation-based models, here coupling with another network might lead to a network to become more robust against failures.} To the best of our knowledge, the only other model where coupling can improve robustness is studied by Brummitt et al. \cite{brummitt2012suppressing}, which constitutes an extension of the sandpile model.  
Perhaps more interestingly, we also see that
the optimal robustness (i.e., largest $p_{\star}$) is attained at a non-trivial coupling level $a = \simeq 0.53$. This suggests that coupling has a multi-faceted impact on robustness and that systems are most robust when they are coupled in a specific, non-trivial way; in Section \ref{subsec:optimization} we provide some concrete ways to identify such optimum coupling levels.


In addition to affecting the system robustness in non-trivial ways, we  see from Figure \ref{fig:2_net_trans} that changing the coupling level can also give rise to different (and, sometimes very interesting) transition behaviors. In particular, we see that network $A$ can go through any one of the transitions demonstrated in previous work \cite{yingrui2016cascadingfailure,yaugan2015robustness} for single networks (see Figure \ref{fig:single_trans}) depending on its coupling level with network $B$. More interestingly, when coupled to network $B$ at a specific level, i.e., with $a=b=0.37$, it is seen to go through a type of transitions that was not seen in the case when it operates as an isolated network. {\em This behavior can be described as a sequence of first, second, first, second, and first order transitions, and to the best of our knowledge was not seen before in any model} \footnote{We note that the behavior demonstrated here is fundamentally different from the few other cases in the literature where multiple transitions have been reported; e.g., see \cite{multiple_percolation,wu2014multiple}. There, the type or the number of transitions do not change with the level of coupling across the networks. Instead, multiple transitions arise only when networks with {\em different} robustness levels are coupled together, and their total (or, average) size is plotted against the size of the attack that is applied to {\em all} networks involved.}. In this case, the network stabilizes twice after a sudden drop in the network size during the cascading process, before going through an abrupt final breakdown.

\begin{figure}[!t]
        \includegraphics[width=0.5\textwidth]{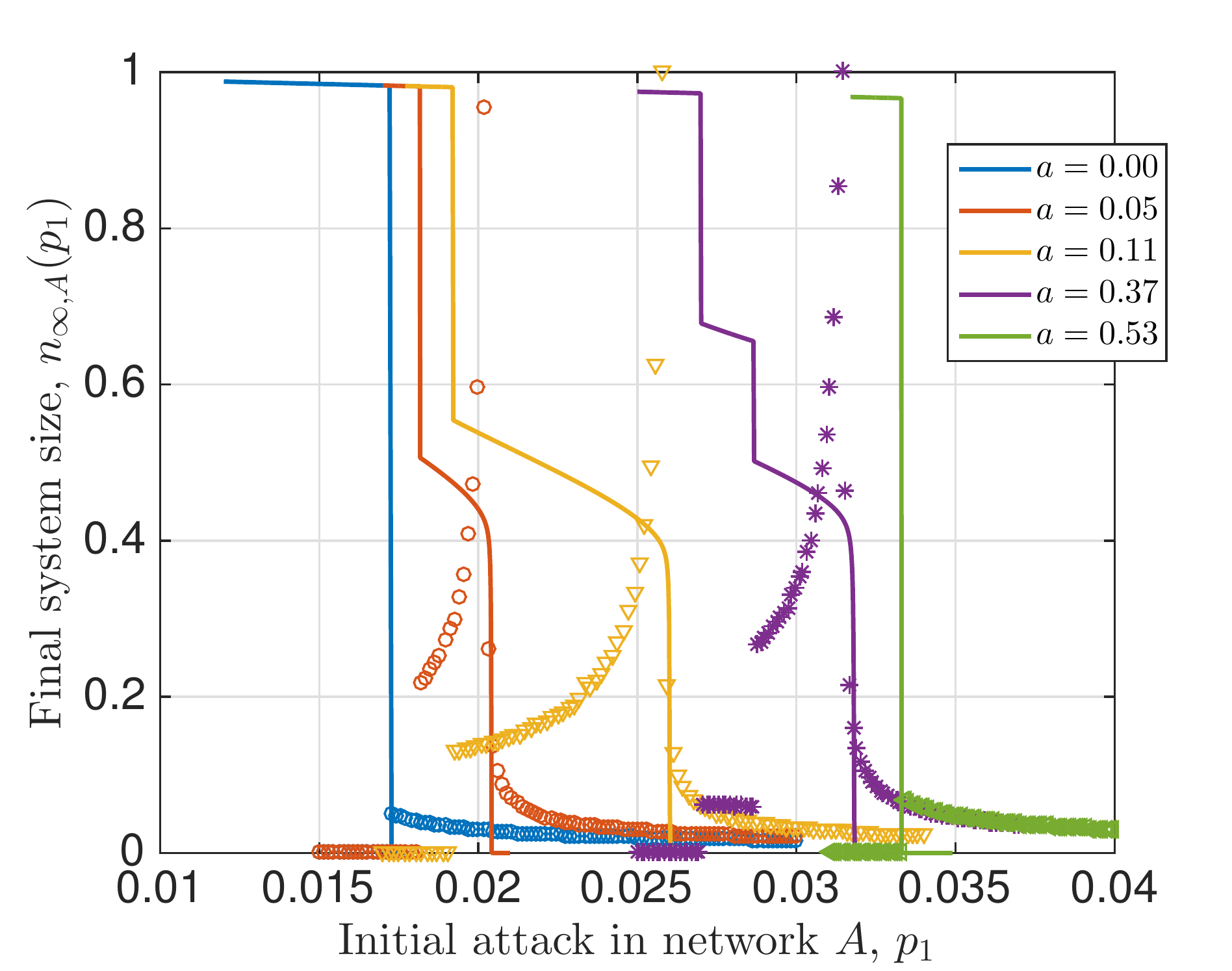}
    \caption{\sl Number of steps needed to reach steady state for identical networks ($a=b$), for various $a$ values. For the case when $a=0.37$, we observe a novel, unforeseen transition behavior.
    }
    \label{fig:noi}
\end{figure}

To further explore the transition behavior during the cascading failure process, we plot the number of iterations (i.e., the number of load redistribution steps) needed for the system to reach steady-state.  The divergence of the number of iterations 
is considered to be a good indicator of the onset
of large failures, and often suggested as a marker of transition points in simulations; e.g., see \cite{pradhan2010failure,parshani2011critical}. We see that this is indeed the case for our model as well. 
In Fig. \ref{fig:noi}, we plot the final system size together with the number of iterations taken to reach that final size. The solid lines represent final system size under different coupling coefficients, and the symbols represent the number of iterations needed (divided by the maximum iterations number, 1000) in each case. We see that the number of iterations needed is piece-wise stable with discontinuous jumps corresponding to the transition points, and it diverges near the final breakdown of the network. In Appendix \ref{sec:appendixA}, we provide a more detailed discussion on the possible correlations between the type and number of transitions a network exhibits with the distribution of its load and free-space.

For a clearer explanation, let us focus on the case when $a=0.37$ (purple asterisks). We see that both discontinuous drops in the final system size coincide with a discontinuous increase in the number of iterations.
As the attack size $p_1$ increases further from that second jump, we see a continuous increase in the number of iterations coinciding with the continuous decrease in final system size. This eventually leads to the number of iterations diverging, and as would be expected coincides with the system breaking down entirely. 



\begin{figure}[t]
\begin{center}
\includegraphics[width=.5\textwidth]{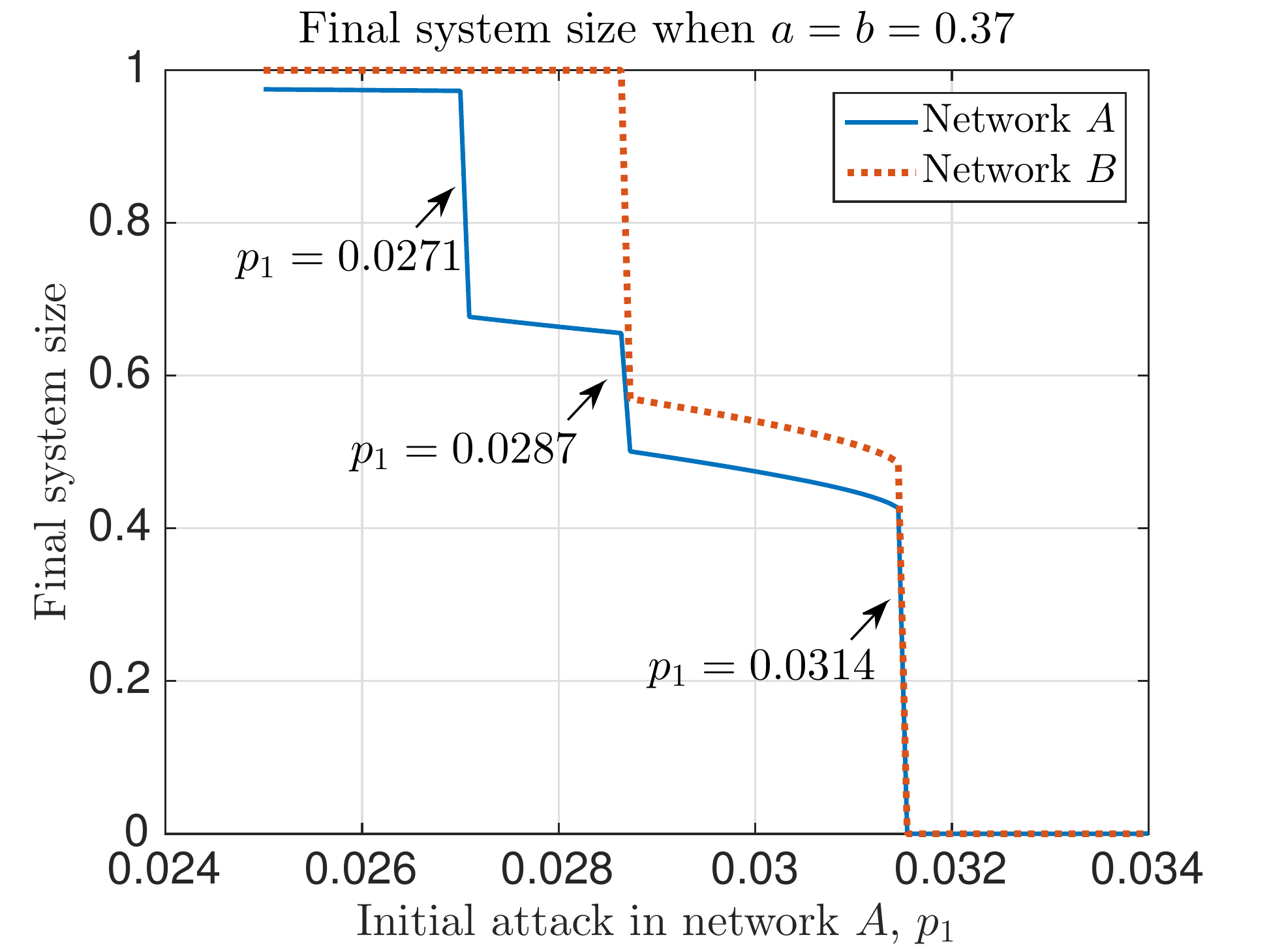} 
\end{center}
\caption{\sl Final system size in two networks when only network $A$ has been attacked initially. The two networks are statistically identical with $a=b=0.36$. Their loads follow a Weibull distribution with $k=0.4$, $\lambda=100$, $L_{min}=10$, and $S=0.6 L$}
\label{fig:final_size_36}
\end{figure}

In Fig.~\ref{fig:final_size_36}, the final system size of network $A$ and $B$ are depicted together (for the case $a=0.37$), showing clearly the effect of interdependence on transition behaviors. Up until $p_1=0.0287$, there are no failed lines in network $B$ although network $A$ already experiences cascading failures; this indicates that all lines in $B$ are able take the extra load from network $A$ even though $A$ loses a significant fraction of its lines at $p_1=0.0271$. When some lines start failing in network $B$ at $p_1=0.0287$, a {\em large} cascade of failures take place causing a significant number of lines fail from both networks marked by discontinuous drop in the final size of both networks. After this point, the remaining system is able to sustain higher initial attacks (because the lines that survive until this point tend to have {\em larger} free-space than average). However, when we reach $p_1=0.0314$, another large cascade takes place that collapses both networks. This final breakdown is observed almost simultaneously in networks $A$ and $B$, primarily because once a network collapses, the other network will need to take over all the load in the system,
and in most cases will not be able survive on its own.

\subsection{Optimizing the robustness of an interdependent system}
\label{subsec:optimization}
Final breakdown point and critical deviation size are good indicators of robustness, but only when we focus on a single network or a specific network in an interdependent system. We now discuss how the robustness of an entire interdependent system can be quantified, with an eye towards identifying {\em optimal} coupling levels that maximize system robustness. Assume that initially $p_1$ fraction of lines from $A$ and $p_2$ fraction of lines from $B$ are attacked randomly. 
The $p_1,p_2 \in [0,1]$ plane is naturally divided into four  \textit {survival regions} \cite{scala2016cascades}.
where 
i) $S_{12}$ represents the initial attack pair $(p_1,p_2)$ under which both networks survive, i.e., have {\em positive} fraction of functional lines when steady state is reached; ii) $S_1$ represents the case where only network $A$ survives; iii) $S_2$ represents the case where only network $B$ survives; and iv) $S_0$ represents the region where no network survives, i.e., the entire system fails with no alive lines. It is then tempting to study the affect of network coupling on these four regions.  

\begin{figure}[t]
\vspace{-3mm}
\centering
\includegraphics[width=0.5\textwidth]{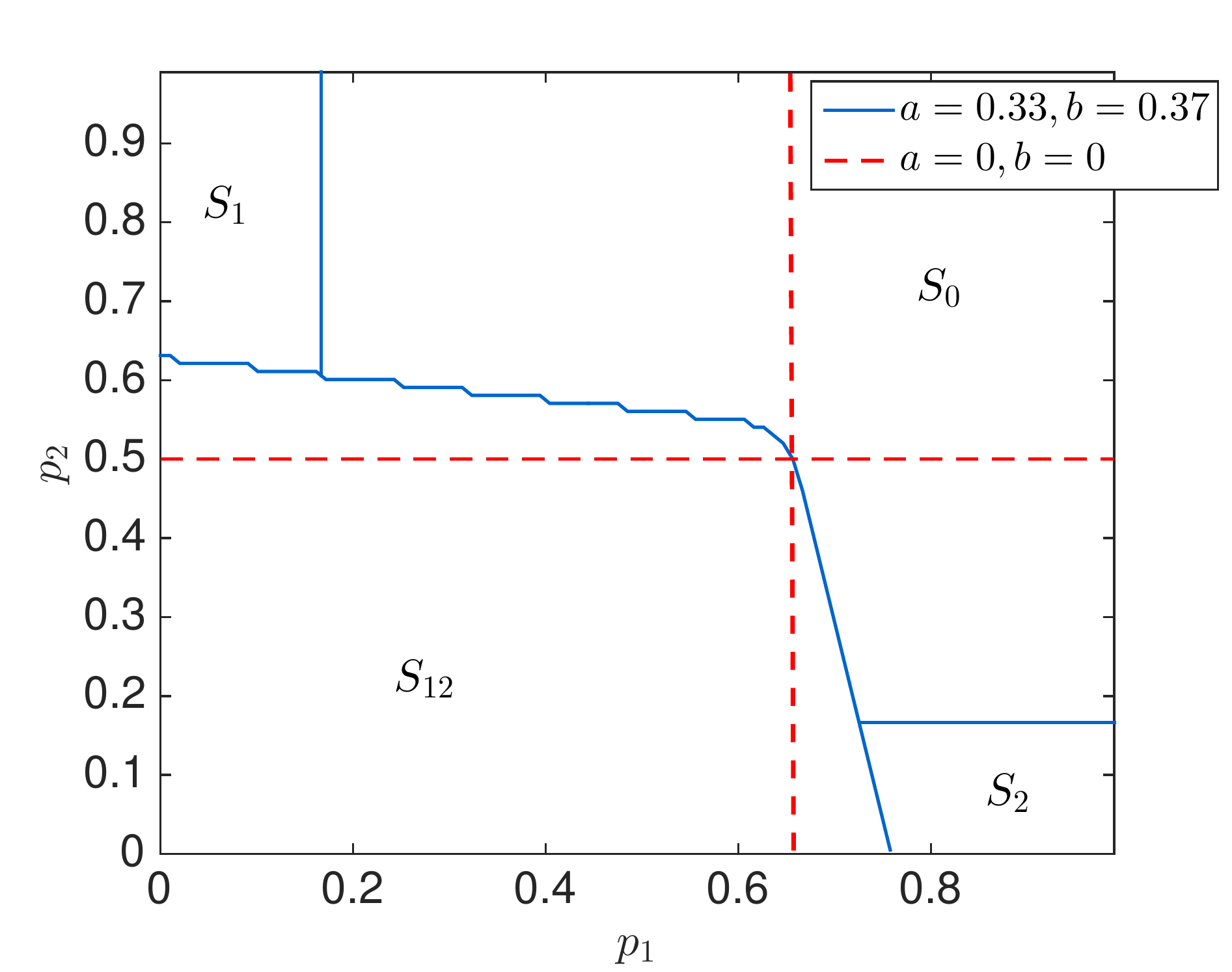}
\caption{\sl Survival regions of the coupled system under load-redistribution based model. When coupling is introduced, regions where both networks survive or collapse ($S_{12}$ and $S_0$, respectively) get larger, while regions where only one network survives ($S_1$ and $S_2$) shrink significantly. 
}
\label{fig:compare_region}
\end{figure}

To provide a concrete example, let network $A$ have $L_A \sim U[10,30]$, $S_A \sim U[40,100]$, and network $B$ have $L_B \sim U[20,40]$, $S_B \sim U[30,85]$, with $U$ denoting {\em uniform} distribution. The initial load distribution and free space distribution are assumed to be independent in each network.
We see from Fig. \ref{fig:compare_region} that when there is no coupling ($a=b=0$), both networks operate in isolation and the survival of $A$ and $B$ are independent from each other; as we would expect, the two dashed lines (in red color) mark the critical attack sizes for $A$ and $B$ when they are in isolation \cite{yingrui2016cascadingfailure}. When we introduce coupling to the system, e.g., with $a=0.33$ and $b=0.37$, we see an interesting phenomenon indicating a multi-faceted impact of coupling on system robustness. The region $S_{12}$ where both networks survives enlarges, while $S_1$, $S_2$ where only one network survives shrink dramatically. Meanwhile, $S_0$ where both networks collapse also enlarges. In a nutshell, when coupled together, the two networks are able to help each other to survive larger attack sizes as compared to the case when they are isolated; however, this comes at the expense of also failing together at smaller attack sizes than before.



\begin{figure}[!t]
\centering
\includegraphics[width=0.5\textwidth]{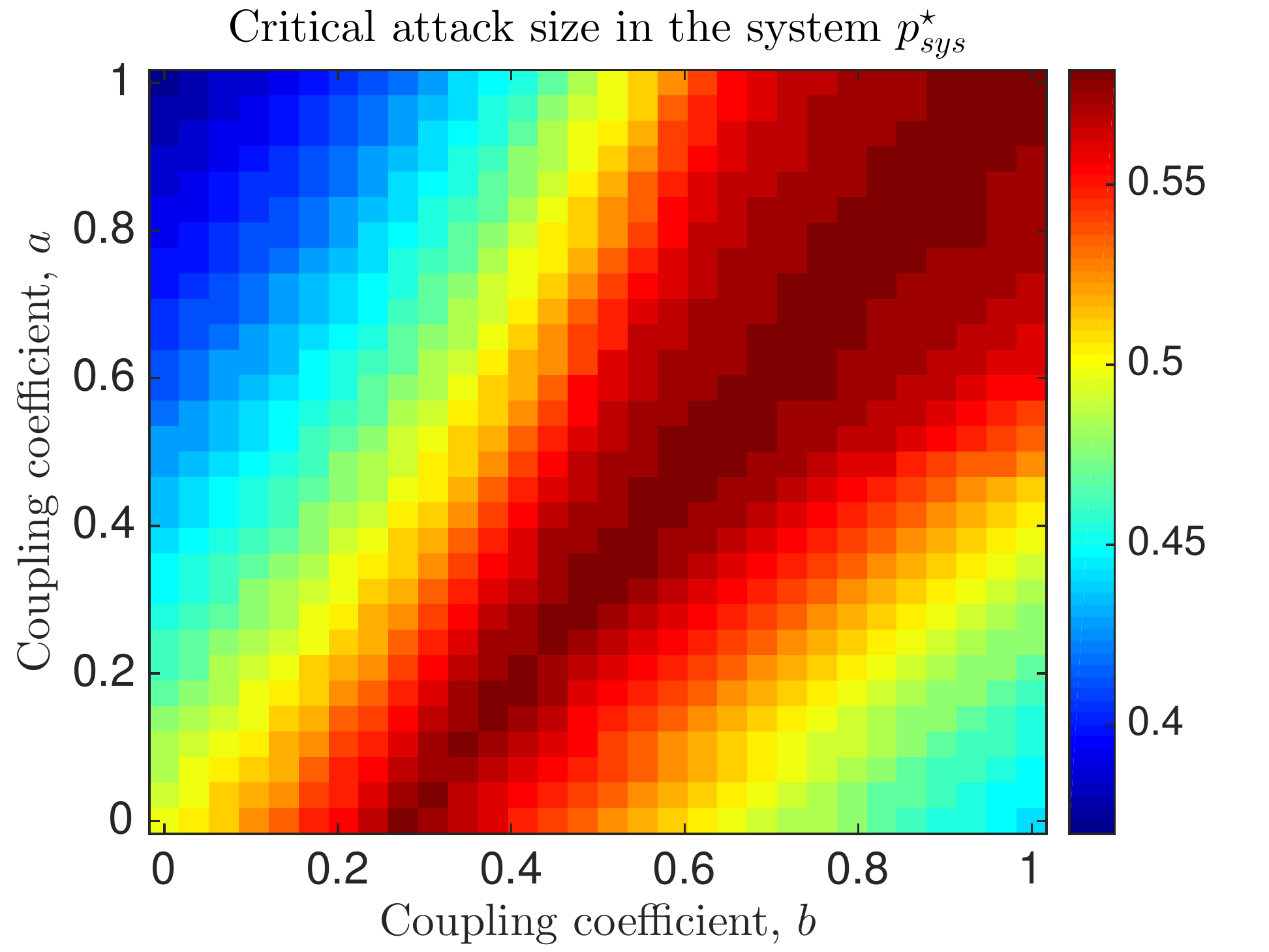}
\caption{\sl Color map of the critical attack size under different coupling coefficients $a$ and $b$. Darker colors indicate larger $p^{\star}_{sys}$ values, meaning that the interdependent system is more robust. 
}
\label{fig:a_pstar}
\end{figure}


To further quantify the effect of coupling on system robustness, we consider the setting above 
while varying the coupling coefficients $a$ and $b$. For both networks, we deploy the same initial attack, i.e., $p_1=p_2=p$, and define the critical system attack size $p^{\star}_{sys}$ as the minimum $p$ that collapses at least one network in the system when cascading failures stop; i.e., $p^{\star}_{sys}$ marks the intersection of the $p_1=p_2$ line and the boundary of the $S_{12}$ region in Figure \ref{fig:compare_region}.

The metric $p^{\star}_{sys}$ proposed here provides a simple and useful way to quantify the robustness of the overall system. For example, aside from being the smallest attack size needed to be launched on both networks to fail at least one of them completely, it gives a good indication of the {\em area} of the $S_{12}$ region where both networks are functional at steady-state. 
In Fig.~\ref{fig:a_pstar} we show the value of $p^{\star}_{sys}$
for different coupling coefficients $(a,b)$ using a color map;
the darker the graph, the larger is the $p^{\star}_{sys}$ value. Using this, one can design an interdependent system to have the optimum coupling levels $(a,b)$ so that robustness of the overall system is maximized (in the sense of maximizing $p^{\star}_{sys}$). We see that the optimum  $(a,b)$ is not unique, but instead contain in a certain {\em strip} of the $[0,1]^2$ plane. This indicates that the robustness of the  interdependent system can be optimized even under certain application-specific constraints on the coupling levels $a$ and $b$; e.g., one might need to have $a=b$ for fairness to both networks, or $a+b=1$ to bound the total load transfer across networks, etc.

\section{Conclusion}
\vspace{-1mm}

\label{sec:conclusion}
In this paper, we studied the robustness of interdependent systems under a flow-redistribution based model. In contrast to percolation-based models that most existing works are based on, our model is suitable for systems carrying a flow (e.g., power systems, road transportation networks), where cascading failures are often triggered by {\em redistribution} of flows leading to {\em overloading} of lines. 
We give a thorough analysis of cascading failures in a system of two interdependent networks initiated by a random attack. 
We show that 
(i) the model captures the real-world phenomenon of unexpected large scale cascades: final collapse is always first-order, but it can be preceded by a sequence of several first and second-order transitions; 
(ii) network robustness tightly depends on the coupling coefficients, and robustness is maximized  at  non-trivial coupling levels in general; (iii) unlike existing models,
interdependence has a multi-faceted impact on system robustness in that interdependency can lead to an  improved robustness for each individual network. 

\section*{Acknowledgement}

This research was supported by the National Science Foundation through grant CCF \# 1422165, and by the Department of ECE at Carnegie Mellon University. A.A. acknowledges financial support by the Spanish government through grant FIS2015-38266, ICREA Academia and James S. McDonnell Foundation.

\bibliography{sample}

\setcounter{equation}{0}
\setcounter{section}{0}
\renewcommand{\thesection}{\Alph{section}.}
\renewcommand{\theequation}{A.\arabic{equation}}

\appendix
\section{Explanation on Multiple Continuous/Discontinuous Transitions}
\label{sec:appendixA}

\begin{figure}[!tbp]
  \centering
  \subfigure[]{\includegraphics[width=0.49\textwidth]{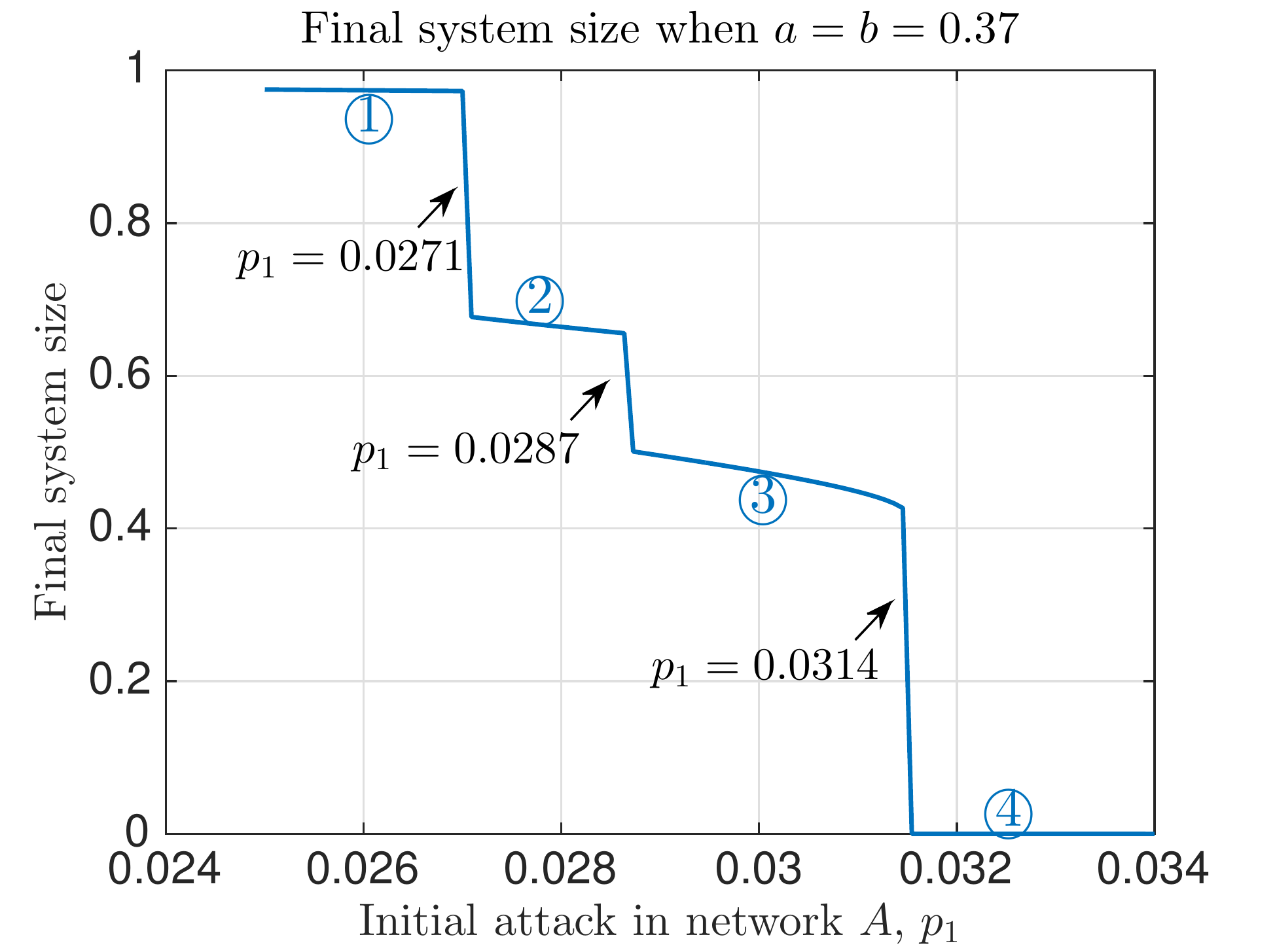}
  \label{fig:q_value_left}}
  \subfigure[]{\includegraphics[width=0.49\textwidth]{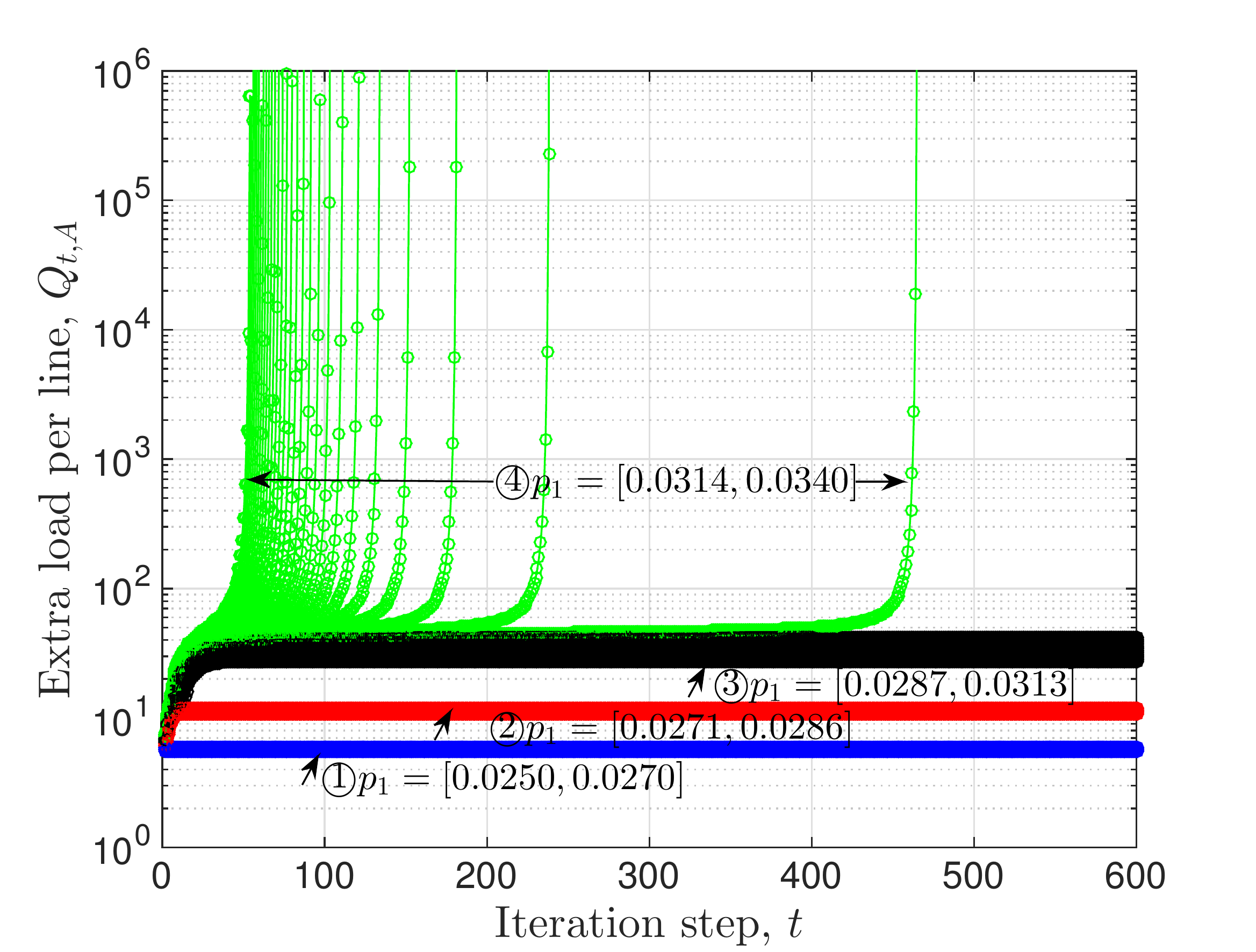}\label{fig:q_value_right}}
  \caption{\sl Extra load per alive line $Q_{t,A}$ is shown (at different attack sizes $p_1$ on Network $A$) as a function of cascade step $t=0,1,\ldots$, for 
  the setting considered in Figure \ref{fig:final_size_36}.  The jumps in the transitions divide the final system curve into four regions (marked with circled numbers), which correspond to four clusters in the $Q_{t,A}$ plots (distinguished by four colors).}
  \label{fig:q_value}
\end{figure}

In this Section, we will explore in more details the underlying reasons for a network to undergo multiple continuous/discontinuous transitions under the flow redistribution model studied in this paper. 
First of all, we note that whether a line survives or fails a particular stage of cascading failure depends on the the {extra load per alive line} at that iteration, i.e., $Q_{t,A}$ or $Q_{t,B}$. With this in mind, in Figure \ref{fig:q_value_right} we plot $Q_{t,A}$ as a function of the iteration step $t$ under the setting of Figure \ref{fig:final_size_36} (i.e., when network $A$ experiences multiple 
transitions). In all cases, we vary attack size $p_1$ over a range with small increments, so that a single curve in Figure  \ref{fig:q_value_right}  represents the change of $Q_{t,A}$ vs. $t$ under a specific attack size $p_1$. 

We observe that each $p_1$ value leads to a variation of $Q_{t,A}$ that belongs to one of the four clusters, distinguished by different colors in Figure  \ref{fig:q_value_right}. For example, as $p_1$ increases from $0.0250$ to $0.0271$, the corresponding $Q_{t,A}$ curves move up smoothly forming the blue cluster. At $p_1=0.0272$, $Q_{t,A}$ experiences a jump, but as $p_1$ increases further,  $Q_{t,A}$ curves move up continuously until $p_1=0.0287$, forming the red cluster. The jump between the blue and red clusters at $p_1=0.0271$ coincides with the first jump in the transition in Figure \ref{fig:q_value_left}. Similarly, at $p_1=0.0287$ we observe a second jump in $Q_{t,A}$ curves between the red and black clusters, which corresponds to the second jump in Figure \ref{fig:q_value_left}. When attack size $p_1$ further increases, $Q_{t,A}$ curves keep moving up smoothly until $p_1=0.0314$ after which $Q_{t,A}$ goes to infinity as $t \to \infty$, meaning that network $A$ collapses completely without any alive lines; the corresponding $Q_{t,A}$ curves for $p_1 \geq 0.0315$ form the fourth cluster show by dotted green lines.
Not surprisingly, $p_1=0.0314$ corresponds to the final breakdown point observed in Figure \ref{fig:q_value_left}. 



Another way to read these figures is that after the extra load per non-failed line $Q_{t,A}$ (resp.~$Q_{t,B}$) reaches a certain value, the network $A$ (resp.~$B$) goes through a sequence of failures after which it either stabilizes with a large fraction of failed lines, or it can not stabilize and goes through a complete breakdown. These {\em critical} values of $Q_{t,A},Q_{t,B}$ and their connection to the emergence of multiple transitions can be understood better in the case of a single network. 
In \cite{yingrui2016cascadingfailure}, we have provided a detailed analysis of the global redistribution model in single networks and demonstrated 
that the critical transition values are determined by the inequality:

\begin{equation} \label{eq:single_ineq}
g(x):=\mathbb{P}[S>x]( x+\mathbb{E}[L \mid S>x]) \geq \frac{\mathbb{E}[L]}{1-p}, \quad x \in (0, \infty)
\end{equation}
With $x^{\star}$ denoting the {smallest solution} of (\ref{eq:single_ineq}), the final system size  is given by 
\begin{equation}
n_{\infty}(p)= (1-p)\bP{S>x^{\star}}.
\label{eq:n_infty}
\end{equation}
Here $x$ represents {\em candidate values} for the extra load per alive line at the steady-state; i.e., it represents potential solutions to $Q_{\infty}$. To see this better,
we can rewrite the inequality (\ref{eq:single_ineq}) as
\begin{equation}
  x  \geq \frac{p \bE{L} + (1-p) \bE{L \1{S \leq x}}}{(1-p)\mathbb{P}[S>x]}.
\label{eq:osy_new_inequality}
\end{equation}
We can now realize that for any $p$ and $x$ for which this inequality holds, the {\em alternative} attack that kills 
i) $p$-fraction of the lines randomly; and ii) all remaining lines whose free-space is less than $x$ (i.e., that satisfy $S \leq x$), is a {\em stable} one that does not lead to any single additional line failure. To see this, note that 
the term $(1-p)\mathbb{P}[S>x] $ in (\ref{eq:osy_new_inequality}) gives the fraction of lines that survive the alternative attack, where each surviving line having at least $x$ amount of free-space, while  $p \bE{L} + (1-p) \bE{L \1{S \leq x}}$ gives the total load failed initially as a result of the alternative attack.
Thus, for a given attack size $p$, the smallest $x$ satisfying  inequality 
(\ref{eq:single_ineq}) or (\ref{eq:osy_new_inequality}) will give us the steady-state extra load per alive line $Q_{\infty}$.

With these in mind, we now explore the underlying reasons for the final system size $n_{\infty}(p)$ to exhibit
(potentially multiple) discontinuous transitions. From Figure \ref{fig:q_value} and the discussion that follows, we expect discontinuous transitions in $n_{\infty}(p)$ to appear simultaneously with discontinuous jumps in the behavior of $Q_t$ as $p$ varies. We now show that our results given at (\ref{eq:single_ineq})-(\ref{eq:n_infty}) confirm this intuition. 
To visualize the implications of (\ref{eq:single_ineq})-(\ref{eq:n_infty}) better, we should  plot $g(x)$ as a function of $x$, and find the leftmost intersection of this curve and the horizontal line drawn at $\frac{\mathbb{E}[L]}{1-p}$. Let this leftmost intersection be denoted by $x^{\star}(p)$ (with the notation making the dependence of $x^{\star}$ on the attack size $p$ explicit). The final system size is given from  (\ref{eq:n_infty}) as $n_{\infty}(p)= (1-p)\bP{S>x^{\star}(p)}$. Assuming that the tail of the distribution of $S$ is continuous, we see that $n_{\infty}(p)$ will exhibit a discontinuous jump if (and only at the points where) $x^{\star}(p)$, which is analogous to  the steady-state extra-load per alive line $Q_{\infty}$, exhibits a discontinuous jump. This confirms the intuition stated above.

\begin{figure}[t]
\vspace{-3mm}
  \centering
  \subfigure[]{\includegraphics[width=0.49\textwidth]{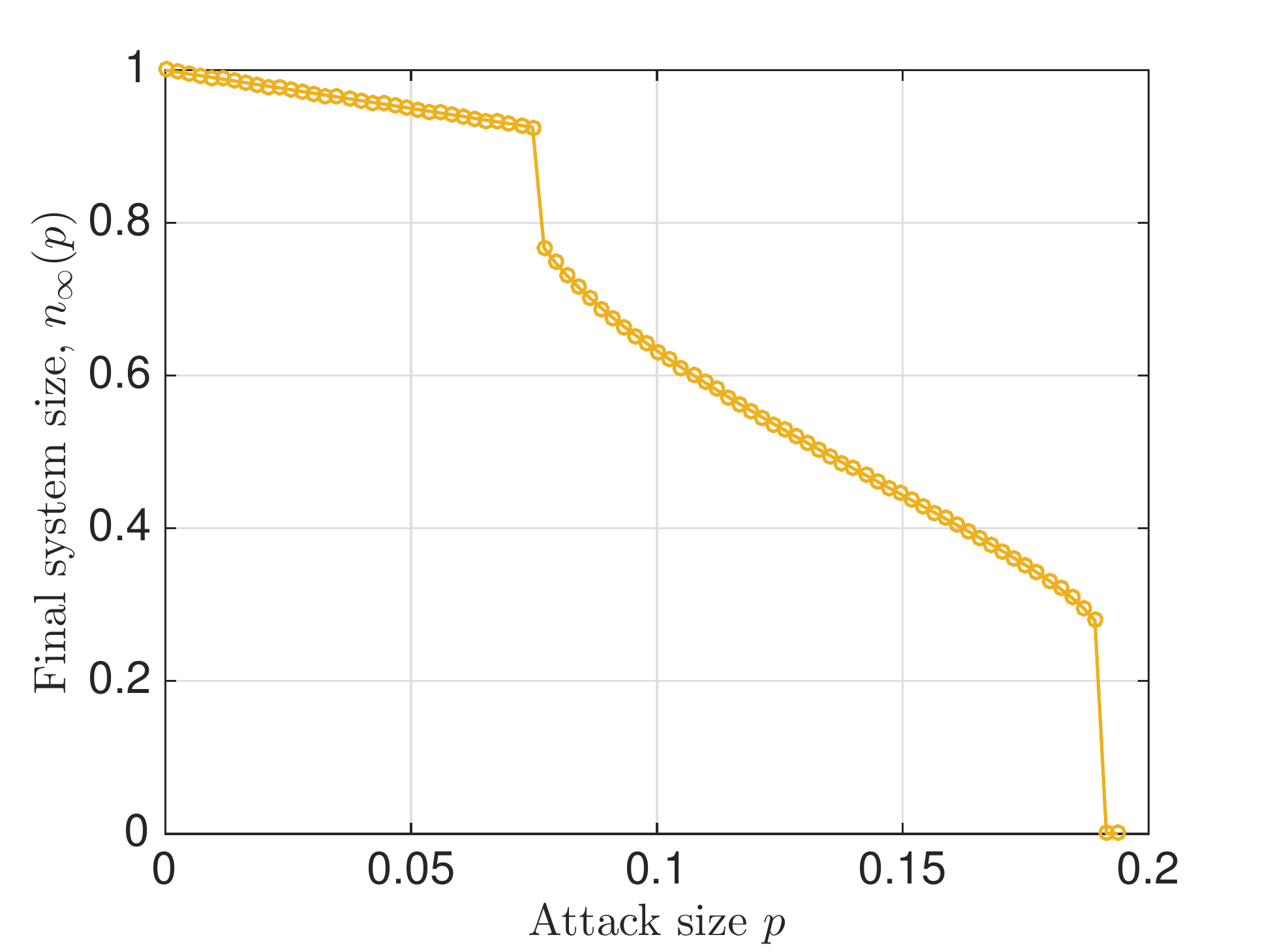}\label{fig:gx_left}}
  \hfill
    \subfigure[]{\includegraphics[width=0.49\textwidth]{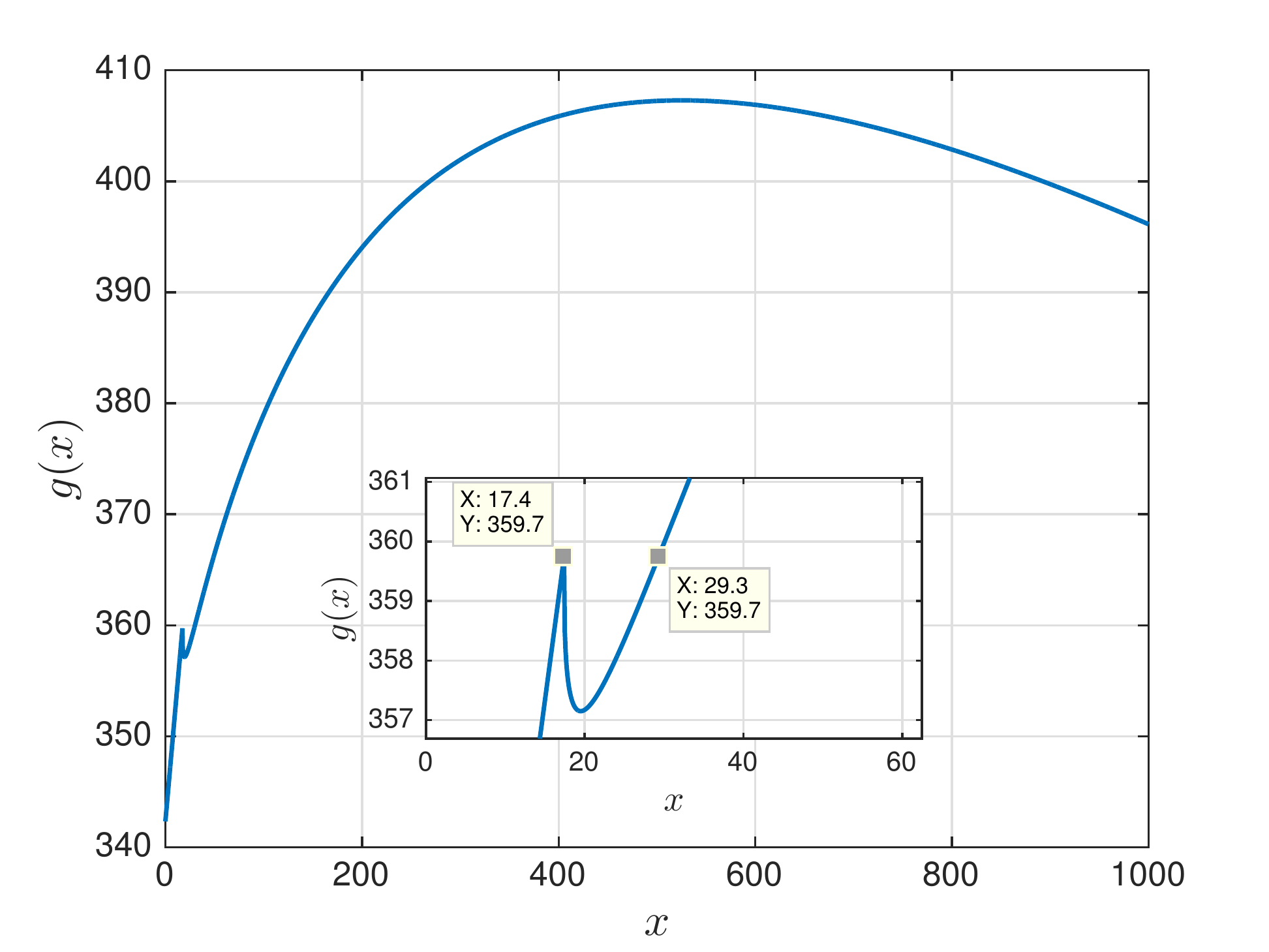}\label{fig:gx_right}}
  \vspace{-2mm} \caption{\sl Multiple transitions in a single network and the corresponding function $g(x)$ (defined at (\ref{eq:single_ineq})) is plotted when $L$ follows Weibull distribution with $k=0.4$, $\lambda=100$, $L_{min}=10$, and $S=\alpha L$ where $\alpha = 1.74$. The Inset zooms in to the region where $g(x)$ has a local maximum.
}
\label{fig:gx}
\end{figure}

Recall that $x^{\star}(p)$ is the leftmost intersection of $g(x)$ and $\bE{L}/(1-p)$, and assume 
 that $\bE{L ~|~ S > x}$
is continuous, so that $g(x)$ is continuous. Then,
$x^{\star}(p)$ (and thus the final system size $n_{\infty}(p)$)
{\em will exhibit one discontinuous jump for every {\em local} and the global maxima of $g(x)$.}
This last statement explains why certain $L,S$ distributions lead only to a single discontinuous jump (since the corresponding $g(x)$ has a single maxima) while others give two (or, potentially more) discontinuous transitions. 
An example for the latter case is given in Figure \ref{fig:gx}. We see that 
the corresponding function
$g(x)$ (Figure \ref{fig:gx_right}) exhibits a {\em local} maxima at $x=17.4$. As a result, when we search for the leftmost intersection of  $g(x)$ and $\bE{L}/(1-p)$ as $p$ varies from zero to one, we see that at a certain $p$ value, the leftmost solution $x^{\star}(p)$  jumps 
from $x=17.4$ to $x=29.3$, creating a first-order transition in the final system size  $n_{\infty}(p)= (1-p)\bP{S>x^{\star}(p)}$. After this point, as $p$ increases further, the (leftmost) intersection points increase smoothly, leading to the continuous transition seen in Figure \ref{fig:gx_left},  until the global maxima of $g(x)$ is reached. At that $p$ value, the leftmost intersection of $g(x)$ and $\bE{L}/(1-p)$ jumps from a finite value to {\em infinity} (indicating that there is no $x$ satisfying  inequality (\ref{eq:single_ineq})), and the system goes through a discontinuous transition leading to its complete break down. 

\section{Simulation Results under Global-Local Combined Redistribution Model}

The main problem considered in this paper, concerning the cascade of failures in two interdependent flow networks, would be expected to depend on the network connectivity patterns in practical scenarios. However, the approach used in this paper offers physical insight by 
proposing a {\em mean field} approach on the setup presented.
In fact, the abstraction used in this paper is equivalent {\em in spirit} to the determination of percolation properties based on degree distributions, mean-field, heterogeneous mean-field, and generating function approaches, etc. In addition, merely topology-based models  where the failed load is redistributed solely in the local neighborhood of the failed line (e.g., as in \cite{MotterLai,wang2008universal,mirzasoleiman2011cascaded}) suffers from two main issues. First of all, it is often not possible to obtain complete analytic results under topology-based redistribution models, even within the single network framework. Thus, unlike the detailed analytical results given in this paper for interdependent networks, one would most likely be constrained to simulation results 
if a topology-based redistribution model was used. Secondly, 
 models where the failed flow gets redistributed only locally according to a topology cannot capture the {\em long-range behavior} of failures that are observed in  most real-world cascades \cite{scala2016cascades}. 
 
 With these in mind, we believe our paper exercises a reasonable trade-off of capturing key aspects of real-world cascades while being able to obtain complete analytic results. Nevertheless, we find it useful to complement our analytical results with simulations that demonstrate how network topology affects the robustness properties of interdependent networks. To this end, we consider a model that combines the global redistribution model described in Section \ref{sec:model}  and the local redistribution model used in \cite{MotterLai}. In particular, assume that upon failures in a network, a $\gamma$-fraction of the failed flow is redistributed solely in the local neighborhood of the failed line, while the rest gets redistributed among {\em all} functional lines. In the case of interdependent networks studied here, we only focus on the intra-topology of networks $A$ and $B$ and still couple them through parameters $a$ and $b$; i.e., when a line in $A$ fails, $a$-fraction of the failed flow gets redistributed equally among {\em all} functional lines of $B$, while  $(1-a)\gamma$-fraction gets redistributed locally in $A$ among the neighbors of the failed line, and the remaining $(1-a)(1-\gamma)$-fraction gets redistributed among {\em all} functional lines of $A$. 
 
\begin{figure}[!t]
\centering
\includegraphics[width=0.49\textwidth]{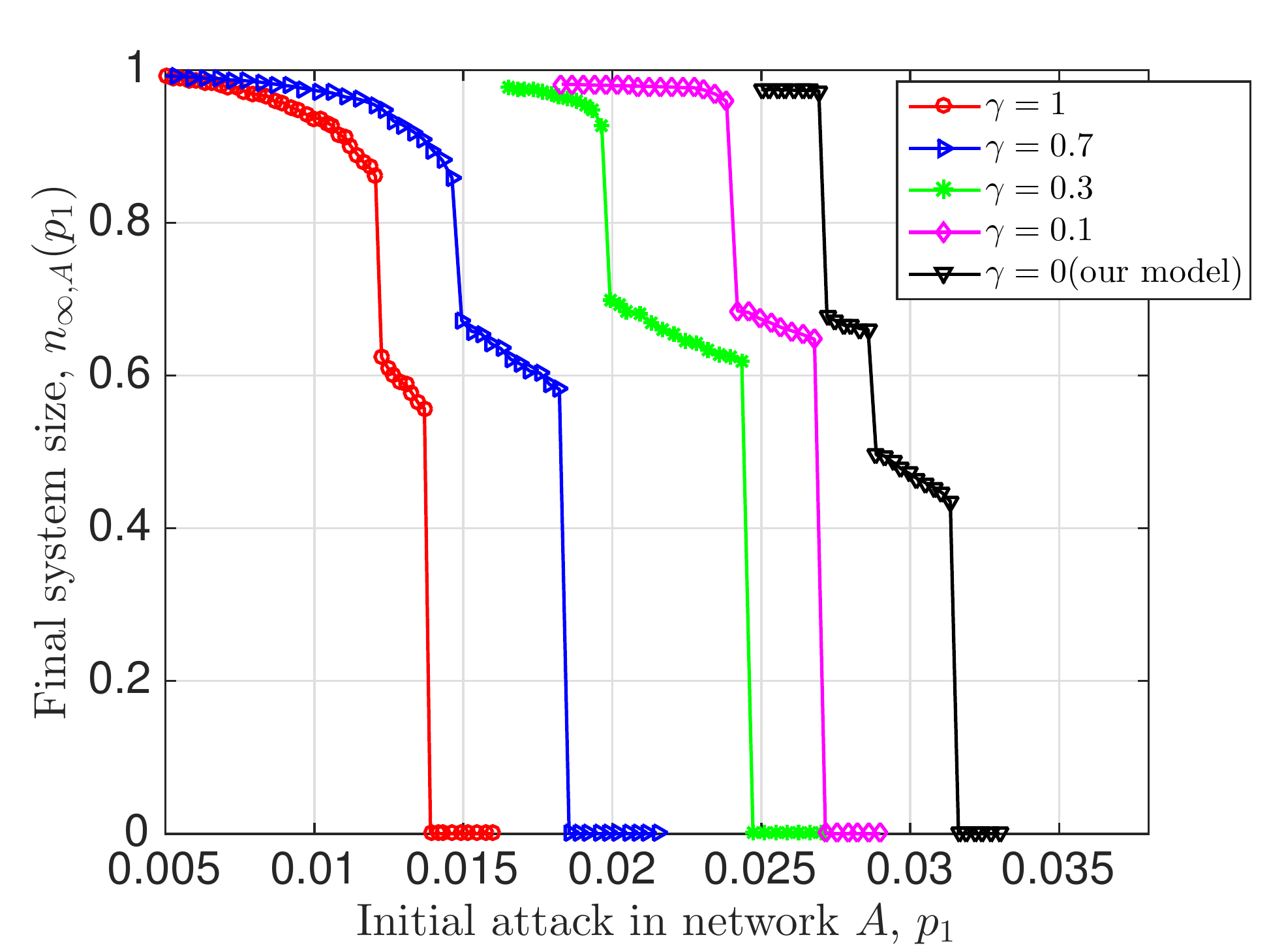}
\vspace{-2mm} \caption{\sl  Effect of  parameter $\gamma$, which controls the fraction of failed load that will be redistributed locally according to network topology, on the robustness of interdependent systems. 
}\label{fig:compare_gamma}
\end{figure}

 With this approach, we recover the model analyzed in our paper when $\gamma=0$, while setting $\gamma=1$ gives a merely topology-based model. We now present a simulation result that shows the robustness of an interdependent system under different $\gamma$ values. For convenience, we consider the same set-up used in Fig.~\ref{fig:final_size_36}, i.e. the two networks are statistically identical with coupling coefficient $a=b=0.36$, and their loads follow a Weibull distribution with $k=0.4$, $\lambda=100$, $L_{\textrm{min}}=10$, and $S=0.6L$. For simplicity, we assume that the topologies of both networks are generated by the Erd\H{o}s-R\'{e}nyi model with $9000$ nodes and link probability 0.2, leading to a mean number $N$ of links around $8.1 \times 10^6$. 

 The results are depicted in Figure \ref{fig:compare_gamma}. As would be expected, as $\gamma$ decreases from one (purely topology-based model) to zero (the model analyzed in our paper), the robustness of network $A$ increases. In other words, the more fraction of failed flow gets shared globally instead of locally, the more robust the network becomes. This is intuitive since when failed flow is shared globally, the additional load per functional line decreases, leading to a lower chance of triggering cascading failures. Nevertheless, the qualitative behavior of the robustness of network $A$ as the attack size $p_1$ increases remains relatively unchanged at different $\gamma$ values; e.g., in all cases, we observe multiple discontinuous transitions, with continuous transitions in between. This suggests that the mean-field approach used in our analysis (i.e., the case with ($\gamma=0$))
 is able to capture very well the {\em qualitative} behavior of final system size for all $\gamma$ values.
 

\end{document}